\documentclass[
aps,prd,nofootinbib,superscriptaddress,twoside,twocolumn,nofootinbib,showpacs
]{revtex4-2}

\usepackage{amsmath}
\usepackage{graphicx,float}
\usepackage{amssymb}
\usepackage{subfig}
\graphicspath{{figures/}}
\usepackage[colorlinks=true,linkcolor=blue,urlcolor=blue,citecolor=blue]{hyperref}
\usepackage{cleveref}
\hypersetup{colorlinks=true,citecolor=blue,linkcolor=blue,urlcolor=blue}

\begin{document}
	
	\title{ Quantum orientation entanglement analysis of the interpolating helicity states between the instant form dynamics and the light-front dynamics }
	\author{Deepasika Dayananda and Chueng-Ryong Ji}
	\affiliation{Department of Physics and Astronomy, North Carolina State University,  Raleigh, North Carolina 27695-8202, USA}
	
	\begin{abstract}

    The interplay between quantum orientation entanglement and Wigner rotation plays a fundamental role in understanding the behavior of spin angular momentum in quantum states. To analyze the quantum orientation entanglement of the relativistic helicity states interpolating between the Jacob-Wick helicity and the light-front helicity, we examine the relative angle between the particle’s momentum direction and the spin orientation for the interpolating helicity states. For this analysis, we introduce a novel method for expanding the interpolating helicity states in terms of the Jacob–Wick helicity. The corresponding probabilistic coefficients follow the structure of the Wigner d-matrix elements, which we use for the interpretation of the quantum orientation entanglement manifested in the angular distributions of the interpolating scattering helicity amplitudes. As an explicit demonstration, we compute the interpolating helicity amplitudes for the pair production of spin-1 (vector) particles in the annihilation of two spin-0 (scalar) particles, focusing primarily on their contact interaction. In particular, we identify the critical interpolation angle that bifurcates the dynamical branches between the instant-form dynamics and the light-front dynamics and discuss the underlying orientation entanglement in the interpolating helicity amplitudes.

	\end{abstract}
	\maketitle

	\section{Introduction}
	\label{Sec:I}
	
Since the spin-$1/2$ nature of the electron was experimentally observed~\cite{Gerlach1922, Uhlenbeck1926}, the debate on the conceptual issues of quantum entanglement has progressed to establish the current understanding of the quantum nature of matters and applying it to realize quantum computing~\cite{Feynman1982} and quantum information technologies~\cite{QuantumCryptography, Quantumteleporting}. Historically, the debate on the EPR paradox~\cite{PhysRev.47.777} and the complementarity~\cite{PhysRev.48.696} played the important role to formulate the Bell's inequality~\cite{bell1964einstein} which led to its experimental tests~\cite{Aspect82, Clauser72, Zeilinger98} pioneering quantum information science and establish the current stage of quantum physics.

Another reality of nature on par with the quantum entanglement is the relativistic spacetime causality, which defies the concept of the absolute time and unbounded speed of transferring information ~\cite{Einstein1905}. The fundamental framework to accommodate both the quantum entanglement and the relativistic spacetime causality is provided by the relativistic quantum field theory (RQFT) ~\cite{Dirac1928,Schweber1962}. 
Although the non-relativistic quantum mechanics inherits the fundamental concept of quantum entanglement, it lacks the reality of the relativistic spacetime causality due to the notion of the absolute time which leads to the Galilean spacetime symmetry~\cite{LevyLeblond1963}. The most dramatic difference between the non-relativistic Galilean transformation and the relativistic Lorentzian transformation may be realized in the boost operation, namely the non-relativistic boost operations do commute each other for any directions while the relativistic boost operations in different directions do not commute but generate the angular momentum.  
In relativistic dynamics, for example, two subsequent transverse boosts generate a rotation around the longitudinal axis, which yields the well-known Thomas precession phenomenon  \cite{Thomas1926} that corrects the factor of the spin-orbit coupling in the atomic system.
It is interesting to note that the relativistic boost operations can be associated with the rotation while no such mixing between the boost and rotation can occur in non-relativistic dynamics. In this respect, it is important to distinguish the understanding of the quantum entanglement phenomena formulated in non-relativistic dynamics vs. relativistic dynamics.  
Although the spin as an example is well defined both in non-relativistic and relativistic dynamics, one should promote the notion of the spin to the helicity which involves the effect of boosting and/or rotating the spin in order to properly analyze the polarization phenomena that typically characterize the nature of quantum entanglement. To discuss the realization of both quantum entanglement and relativistic causality effects together, we 
present the study of relativistic helicity formulation in this work exemplifying specific computation of scattering helicity amplitudes.
 
For the study of RQFT, Dirac proposed three different forms of the relativistic Hamiltonian dynamics in 1949 \cite{RevModPhys.21.392}: i.e., the instant ($t = 0$), front ($\tau  = (t+z)/ \sqrt{2} =0$), and point ($x^ \mu x_\mu=a^2 > 0, x^0 > 0$) forms. The instant form dynamics (IFD) of quantum field theories is based on the usual equal time quantization, which provides a traditional approach evolved from the non-relativistic dynamics and closely related to the Euclidean space. The quantization with the light-front time yields the light-front form dynamic (LFD), which mainly works with the Minkowski space and
provides the maximum number (seven) of kinematic generators in the Poincar\'e group.
While the quantization in the point form has been used in string theory and conformal field theories~\cite{Point1973,Point1998,Point2001,Point2007}, the IFD and the LFD appear the most popular choices in the discussion of RQFT. In particular, both IFD and LFD offer their own advantages and perspectives in formulating the helicity to realize the quantum entanglement with the relativistic causality. In this work, we thus correspond the realization of the quantum entanglement to each other between the IFD and the LFD with the interpolation methodology that   
we have pursued perviously~\cite{PhysRevD.64.085013,PhysRevD.87.065015, PhysRevD.91.065020, PhysRevD.92.105014, PhysRevD.98.036017, PhysRevD.104.036004}. 

To link the IFD with the LFD, we introduced a parameter $\delta$ in transforming the spacetime coordinates, $x^{\hat{\mu}}$ = $G^{\hat{\mu}}$$_\nu $$x^\nu $, i.e.

	\begin{equation}
    \label{InterCorU}
	\begin{pmatrix}
	x^{\hat{+}}\\
	x^{\hat{1}}\\
	x^{\hat{2}}\\
	x^{\hat{-}}\\
	\end{pmatrix}
	=
	\begin{pmatrix}
	\cos{\delta} & 0 & 0 & \sin{\delta}\\
	0 & 1 & 0 & 0 \\
	0 & 0 & 1 & 0 \\
	\sin{\delta} & 0 & 0 & -\cos{\delta}	\\
	\end{pmatrix}
	\begin{pmatrix}
	x^0\\
	x^1\\
	x^2\\
	x^3\\
	\end{pmatrix} ,
	\end{equation}
where ``hat" ($ ``\string^ "$)	on the indices has been used to denote the interpolating variable with the parameter $0 \leq \delta \leq \pi/4 $.
For the limit $ \delta \rightarrow  0$  usual time coordinates have been recovered with $z$ - axis is inverted ($x^{\hat{+}}=x^0 $\,,\,$x^{\hat{1}}=x^1$\,,\,$x^{\hat{2}}=x^2$\,,\,$x^{\hat{-}}=-x^3$) while for the limit $ \delta \rightarrow \pi/4 $ leads to the standard light front coordinates ($x^{\hat{\pm}}=(x^0 \pm x^3)/ \sqrt{2} = x^\pm  $\,,\,$x^{\hat{1}}=x^1$\,,\,$x^{\hat{2}}=x^2$\,). The same transformation also applies to momentum as well. Since the perpendicular components remain the same, from now on we will omit the ``hat" ($ ``\string^ "$) notation for the perpendicular indices in a four vector.

While the relativsitic helicity is typically defined in the IFD ($\delta = 0)$, known as the Jacob-Wick helicity \cite{JACOB1959404}, the helicity defined in the LFD ($\delta = \pi/4$) is characteristically different from it and thus called as the light-front helicity~\cite{PhysRevD.67.116002}. In particular, it has been noted that the actual spin of the light-front helicity ``+1" points to the opposite direction of the spin pointing in the Jacob-Wick helicity ``+1" when the particle is moving in the negative $z$-direction\cite{Xiji97, RADYUSHKIN1996417}. Therefore, it is crucial to explicitly understand the relative spin and momentum changes taking place in both dynamics with respect to their definitions to avoid confusions when considering the orientation entanglement of helicity. In this work, we
discuss both the Jacob-Wick helicity in IFD and the light-front helicity in LFD and elaborate the interpolating helicity between IFD and LFD with the parameter $\delta$
($0 \le \delta \le \pi/4$). To explore the orientation entanglement with the relativistic causality in place, we exemplify the computation of the scattering helicity amplitudes for the entire region
$0 \le \delta \le \pi/4$. 
While we discuss the interpolating helicity amplitudes in the scattering processes, we focus on their angular distributions to manifest the orientation entanglement of interpolating helicity states. We present our findings of effective description 
in regard to the orientation entanglement.
 
The outline of our work is as follows.
In Sec.~\ref{Sec:II}, we discuss
the quantum orientation entanglement of spin systems stemmed from the fundamental entanglement of spin-1/2, namely, the rotation of $720^0 $  angle to get back to the original phase of the spin-1/2 particle. We motivate our work from the well-known orientation entanglement of spherical harmonics under the the $180^0$ rotation (or reflection) given by $Y_1^0(\theta+\pi)=-Y_1^0(\theta)$ for the later discussion of the quantum entanglement effect in the helicity amplitude of the scattering process between composite particles of spin-1/2 such as the spin-triplet (vector) and spin-singlet (scalar) particles.  
In Sec.~\ref{Sec:III}, we discuss the relativistic helicity formulations starting from the Jacob-Wick helicity in IFD and motivating an alternative helicity formulation, in particular, the light-front helicity in LFD. In Sec.~\ref{Sec:IV}, we present the explicit helicity spinor representations of IFD and LFD and interpolate them with the parameter $\delta$ and summarize the consequences of the spin orientation with respect to the momentum direction of the particle carrying the spin (Sec.~\ref{Sec:IV.A}). 
Then, we express the interpolating helicity in terms of the Jacob-Wick helicity and find the relationship between the interpolating helicity and the Jacob-Wick helicity which turns out to be rather simply given by the Wigner d-functions dependent on the interpolation angle $\delta$ (Sec.~\ref{Sec:IV.B}). 
In Sec.~\ref{Sec:V}, we apply our findings in analyzing the explicit scattering process $SS \to VV$, which involves the
computation of the scattering helicity amplitudes for the 
production of
two spin-one (``Vector" V) particles in the annihilation of two spin-zero (``Scalar" S) particles. To illustrate the main points in the simplest but still highly nontrivial scattering amplitude, we focus on the 
Seagull channel without involving any complication from the non-zero impact parameter in the scattering process and analyze each and every independent scattering helicity amplitudes: $M_\delta^{++}=M_\delta^{--}$ 
(\ref{Amp++}), $M_\delta^{+-}=M_\delta^{-+}$ 
(\ref{Amp+-}),
$M_\delta^{+0}=-M_\delta^{-0}=M_\delta^{0+}(\theta+\pi)=-M_\delta^{0-}(\theta+\pi$) 
(\ref{Amp+0}) and $M_\delta^{00}$ 
(\ref{Amp00}). Summary and conclusions follow in Sec.~\ref{Sec:VI}. 

We added several appendices for completeness. 
In Appendix~\ref{App: A}, 
we summarize the interpolation space-time metrics $g^{ \hat{\mu}\hat{\nu} }$ and the relationship between interpolating covariant and contravariant indices, along with some other important interpolating relations. 
In Appendices~\ref{App: B} and ~\ref{App: C}, we discuss the orientation entanglement of interpolating spin-1/2 and spin-1 spinors, respectively. 
Non-relativistic helicity amplitudes of the Seagull channel and a discussion of the relativistic effect of the longitudinal polarization vector in the helicity amplitudes are presented in Appendices~\ref{App: D} and \ref{App: E}, respectively.
	\section{Orientation entanglement of Spin}
    	\label{Sec:II}
        
In quantum mechanics, the tensor product of two state spaces can be written as the direct sum of the representations of total spin angular momenta. The well-known example is writing the tensor product of two fundamental spin-1/2 representations decomposes into a direct sum of a spin-1 (triplet) and a spin-0 (singlet) representation. Thus, this forms a four-dimensional Hilbert space;
\begin{equation}
\frac{1}{2}\otimes \frac{1}{2} = 1\oplus 0 .
\end{equation} 
Explicitly, one can organize the base states according to the eigenvalues($j$ and $m$). For triplet states, the wave functions $(j=1,m=1,0,-1)$ are symmetric as exemplified by 
\begin{equation}
\begin{split}
|1,1>&= |\uparrow> \otimes |\uparrow >,\\
|1,0>&=\frac{1}{\sqrt{2}} \big(|\uparrow> \otimes |\downarrow >+ |\downarrow> \otimes|\uparrow >\big),\\ 
|1,-1>&=|\downarrow> \otimes | \downarrow >.
\end{split}
\end{equation}
For the singlet state, the wave function $(j=0,m=0)$ is anti-symmetric as exemplified by
\begin{equation}
\begin{split}
|0,0>&=\frac{1}{\sqrt{2}} \big(|\uparrow> \otimes |\downarrow >-|\downarrow> \otimes|\uparrow >\big).\\
\end{split}
\end{equation}
After identifying the entangled states of spin-1/2, we can focus on their spin orientations. In general, the spin orientation entanglement can be easily derived using the most familiar Wigner-d transformation, which shows the quantum states under rotations about a fixed axis. Specifically, it provides the matrix elements of an active rotation about the y-axis by an angle  $\theta_s$; $R_y ^j (\theta_s) =e^{ -i J_y  \theta_s}$. After applying the rotation, the rotated spin-1/2 states can be   expressed as a superposition of the basis state vectors of spin-1/2, namely \(|\uparrow\rangle\) and \(|\downarrow\rangle\),i.e.	
	\begin{subequations}
		\begin{equation}
		\label{spin 1/2 up rotation}
		\begin{split}
		 R_y^{1/2} (\theta_s)  |\uparrow>= &\cos\big(\frac{ \theta_s}{2}\big)|\uparrow> + \sin\big(\frac{ \theta_s}{2}\big) |\downarrow>,
		\end{split}
		\end{equation}
		\begin{equation}
		\label{spin 1/2 down rotation}
		\begin{split}
	 R_y^{1/2} (\theta_s) |\downarrow>=& -\sin\big(\frac{ \theta_s}{2}\big)|\uparrow>+ \cos\big(\frac{ \theta_s}{2}\big) |\downarrow>.
		\end{split}
		\end{equation}
\end{subequations}
The rotation of $180^0$ makes $|\uparrow> \,\ \rightarrow |\downarrow >$, but $|\downarrow > \,\ \rightarrow -|\uparrow >$, revealing the orientation entanglement nature of the spin-1/2 particles. Only after the $720^0$ of rotation, the spin-1/2 states can return to their original configuration. 

The probabilistic coefficients of these states (Eqs. (\ref{spin 1/2 up rotation}) and (\ref{spin 1/2 down rotation})) are given by the usual  Wigner-d matrix elements: $d_{m',m} ^{(j)} ( \theta_s) $ which represents the amplitude for the rotated state to be found in $|j, m' >$ when the original unrotated state is given by  $|j,m> $.
The tensor product of the fundamental spin-1/2 rotated states given in  Eqs. (\ref{spin 1/2 up rotation}) and (\ref{spin 1/2 down rotation}) can be used to derive the rotated states of the triplet ($j=1$) and singlet ($j=0$) 
states; 
\begin{equation}
\begin{split}
\label{TenRo}
R_y^1 (\theta_s)|1,1>= &R_y ^{1/2} (  \theta_s) |\uparrow> \otimes \,\ R_y ^{1/2} (  \theta_s) \uparrow >, \\
R_y^1 (  \theta_s)|1,0>=&\frac{1}{\sqrt{2}} \big( R_y ^{1/2} (  \theta_s)|\uparrow>  \otimes R_y ^{1/2} (  \theta_s)|\downarrow >\\ &+R_y ^{1/2} (  \theta_s) |\downarrow>  \otimes \,\  R_y ^{1/2} (  \theta_s) |\uparrow >\big), \\
R_y^1 (  \theta_s)|1,-1>&=R_y ^{1/2} (  \theta_s) |\downarrow>  \otimes \,\ R_y ^{1/2} (  \theta_s) |\downarrow >,\\
R_y^0 (  \theta_s)|0,0>=&\frac{1}{\sqrt{2}} \big( R_y ^{1/2} (  \theta_s)|\uparrow>  \otimes\,\  R_y ^{1/2} (  \theta_s)|\downarrow >\\ &-R_y ^{1/2} (  \theta_s) |\downarrow>  \otimes \,\  R_y ^{1/2} (  \theta_s) |\uparrow >\big), 
\end{split}
\end{equation}
where the rotation of four-dimensional Hilbert space is expressed in two- dimensional fundamental Hilbert space.
Substituting Eqs.(\ref{spin 1/2 up rotation}) and (\ref{spin 1/2 down rotation}) to Eq.(\ref{TenRo}), one can of course find the 
following relations expressing
the rotated states of four-dimensional Hilbert space as the superposition of themselves;
\begin{equation}
    \label{spin1d}
\begin{split}
R_y^{1} (\theta_s)|1,1 >=& \frac{(1+\cos \theta_s)}{2}|1,1 > \\&+\frac{\sin \theta_s}{\sqrt{2}}|1,0 > +\frac{(1-\cos \theta_s)}{2}|1,-1 >, \\
R_y^{1} (\theta_s)|1,0 >=& - \frac{\sin \theta_s}{\sqrt{2}}|1,1 > \\& + \cos \theta_s |1,0 > +\frac{\sin \theta_s}{\sqrt{2}}|1,-1 >, \\
R_y^{1} (\theta_s)|1,-1 >&=\frac{(1-\cos \theta_s)}{2}|1,1 > \\&  -\frac{\sin \theta_s}{\sqrt{2}}|1,0 > +\frac{(1+\cos \theta_s)}{2}|1,-1 > ,\\
R_y^{0} (\theta_s)|0,0> =&|0,0>.
\end{split}
\end{equation}
These relations in Eq.(\ref{spin1d}) manifest the orientation entanglement of spin-1 states in contrast to the spin-0 state under the rotation of $180^0$ of angle. Namely, under the shift $\theta_s \to \theta_s + \pi$, we note $|1,1>  \rightarrow |1,-1>$ and  $|1,-1>  \rightarrow |1,1>$, but $|1,0>  \rightarrow -|1,0>$ in contrast to $|0,0>  \rightarrow |0,0>$, which distinguishes the ``0" state of spin-1(vector) and spin-0(scalar) particles. The rotation invariance of the spin-0 particle is embedded in its ``scalar" nature while the relations among the spin-1 states under the rotation signify the ``vector" nature represented by the Wigner-d functions $d^{(1)}_{m',m}(\theta_s)$ explicitly given by Eq.(\ref{spin1d}).     

The nature of orientation entanglement in spin-1 systems can be revealed however more dramatically in the system of the two spin-1 particles. For instance, in the back-to-back decay amplitudes to two final vector(spin-1) particles with their respective polarizations from the initial scalar(spin-0) state, the characteristic difference of helicity zero-zero($|1,0>|1,0>$) state vs. helicity plus-minus($|1,1>|1,-1>$) or minus-plus($|1,-1>|1,1>$) state can be identified. Namely,
at the threshold of producing two final vector(spin-1) particles, the produced but rest two final spin-1 states may be identified as one of these three states satisfying the angular momentum conservation, namely $|1,1>|1,-1>$, $|1,-1>|1,1>$ and $|1,0>|1,0>$. The effect of the quantum orientation entanglement appears as the dramatic difference in the
final states. While the magnitude of all three final states is identical to each other, the sign of the amplitude of the final state $|1,0>|1,0>$ is opposite to the amplitude of the other two final states $|1,1>|1,-1>$ and $|1,-1>|1,1>$. We will discuss this remarkable quantum orientation entanglement properties in this work analyzing the relativistic scattering helicity amplitudes of $SS \to VV$, namely the two vector part production process in the annihilation of the two scalar particles.

\section{Non-Relativistic vs.  Relativistic Helicity Formulations}
    	\label{Sec:III}

 In this section, we start from the well-known non-relativistic spin-1 states to discuss the non-relativistic helicity formulation and illustrate its extension to the relativistic helicity formulation, which we will apply for the computation of the scattering process $SS \to VV$. We will use the four-vector notations even in the non-relativistic formulation for the consistency of notations although the transform rules of the four-vectors are different, e.g., Galilean vs. Lorentz transformations in non-relativistic vs. relativistic formulations, respectively.
 
The spin-1 states $|1,m> (m=+1,0,-1)$ can be represented by the polarization four-vectors. In the rest frame of the spin-1 particle, i.e. the four-momentum of particle $p^\mu =(M,0,0,0)$, the polarization vectors corresponding to the spin states $|1,\pm1>$ and $|1,0>$ are respectively given by
	\begin{align}
    \label{PolaRest}
	\epsilon^{\mu}(\pm) &= \mp \frac{1}{\sqrt{2}}(0,1,\pm i,0), \notag \\
    \epsilon^{\mu}(0) &=(0,0,0,1),
	\end{align}
where the polarization vectors are normalized with the use of typical Minkowski metric $g^{\mu\nu}$ 
as $\epsilon(\lambda)\cdot\epsilon^*(\lambda)=g_{\mu\nu}{\epsilon^\mu(\lambda)}{\epsilon^{*\nu}} (\lambda)=-1$ for all $\lambda = \pm, 0$.
The orientation of rest frame polarization vectors can then be written similarly to Eq. (\ref{spin1d}) as
\begin{equation}
\label{spin-rotation}
\begin{split}
R_y^{1} (\theta_s)  \epsilon^{\mu}(+) =& \frac{(1+\cos \theta_s)}{2}\epsilon^{\mu}(+)\\&+\frac{\sin \theta_s}{\sqrt{2}}\epsilon^{\mu}(0) +\frac{(1-\cos \theta_s)}{2}\epsilon^{\mu}(-1), \\
R_y^{1} (\theta_s)\epsilon^{\mu}(0)=& - \frac{\sin \theta_s}{\sqrt{2}}\epsilon^{\mu}(+) \\& + \cos \theta_s \epsilon^{\mu}(0)+\frac{\sin \theta_s}{\sqrt{2}}\epsilon^{\mu}(-1) , \\
R_y^{1} (\theta_s)\epsilon^{\mu}(-1) &=\frac{(1-\cos \theta_s)}{2}\epsilon^{\mu}(+) \\&  -\frac{\sin \theta_s}{\sqrt{2}}\epsilon^{\mu}(0) +\frac{(1+\cos \theta_s)}{2}\epsilon^{\mu}(-1)  .\\
\end{split}
\end{equation}\\
Now, to understand the orientation of the spin states in the moving frames, we need to consider the formulation of the helicity observable. The non-relativistic polarization vectors in moving frame can be derived by transforming the rest frame polarization vectors in Eq.(\ref{PolaRest}), e.g. with the Galilean boost operator in the ${\hat z}$-direction first and subsequently rotate them by an angle \( \theta \) around the \(\hat y\)-axis which is perpendicular to the boosted direction ${\hat z}$. These subsequent operations are denoted as the  transformation $T_g$ given by 
\begin{equation}	  
 	\label{GT}
 T_g=  e^{-i \theta J^2 }e^{-i v K_g^3},
 \end{equation}
 where $J^2$ and $K_g^3$ are the respective generators of the rotation around the $\hat{y}$-axis and the Galilean boost in the $\hat{z}$-direction. Here, the parameters $\theta$ and $v$ correspond to their respective generators $J^2$ and $K_g^3$.
 Applying $T_g$ with the parameters $\theta = \tan^{-1}(\frac{P_1}{P_3})$ and $v=\frac{P_3}{M}$ on the polarization vectors given by Eq.(\ref{PolaRest}) in the rest frame, one can find the polarization vectors for the moving particle of mass $M$ with its four-momentum $p^\mu = (P^0, P^1, 0, P^3)$, noting that $P^0=M$ under the Galilean boost but 
 $P^0=M+\frac{\textbf{P}^2}{2M}$ with the inhomogeneous addition of the kinetic energy in typical non-relativistic kinematics, and the helicity $\lambda = +,-,0$ respectively given by
 \begin{equation}
\begin{split}
\label{GalvecP}
\epsilon^{\mu}_g(P,+)= &- \Big(0, \frac{P^3}{|\textbf{P}|\sqrt{2}},   \frac{ i}{\sqrt{2}},\frac{- P^1}{|\textbf{P}|\sqrt{2}}\Big),\\
\epsilon^{\mu}_g(P,-)= & \Big(0, \frac{P^3}{|\textbf{P}|\sqrt{2}},   -\frac{ i}{\sqrt{2}},\frac{ -P^1}{|\textbf{P}|\sqrt{2}}\Big),\\
\epsilon^{\mu}_g(P,0)=&  \Big(0,  \frac{P^1}{|\textbf{P}|},0,  \frac{P^3}{|\textbf{P}|} \Big),
\end{split}
\end{equation}
where $|\textbf{P}|=\sqrt{(P^1)^2+(P^3)^2}$ and the transverse property $p \cdot \epsilon_g = 0$ of the polarization vector for each and every helicity $\lambda$ is manifest. 
Here, the normalization of polarization vectors $g_{\mu\nu}\epsilon^\mu_g(P,\lambda)\epsilon^{*\nu}_g(P,\lambda)=-1$ is preserved for all $P$ and $\lambda$. 
One should note that the rest frame polarization vectors $\epsilon^\mu(\lambda)$ in Eq.(\ref{PolaRest}) can be retrieved by taking $P^1 = 0$ first and then taking $P^3 = 0$ in reversing the orders of the subsequent operations involved in the transformation $T_g$ given by Eq.(\ref{GT}).    
From these results, we may identify the non-relativistic Galilean helicity operator $\mathcal{J}_g $ as 
	\begin{eqnarray}
    \label{Galilean-Helicity}
	 \mathcal{J}_g 
     &=& T_g J^3 T_g^{-1}\\ \nonumber &=&\frac{\textbf{J}.\textbf{P}}{|\textbf{P}|},
		\end{eqnarray}
where $\textbf{J}.\textbf{P} = J^1 P^1+J^3 P^3$ for the explicit verification.  
Applying this helicity operator in Eq.(\ref{Galilean-Helicity}) to the helicity eigenvectors in 
Eq.(\ref{GalvecP}), we can
assure the recovery of the corresponding helicity eigenvalues, namely $\mathcal{J}_g \epsilon^{\mu}_g(P,\lambda) = \lambda \epsilon^{\mu}_g(P,\lambda)$. 

From the Galilean helicity operator given by Eq.(\ref{Galilean-Helicity}), one can find rather immediately the invariance of the non-relativistic helicity under the rotation but not under the Galilean boost. This can be understood from the difference between the rotation operator $R_{\hat n}(\theta)$ which rotates the state around ${\hat n}$-axis by an angle $\theta$ and the Galilean boost operator $G_{\hat n}(v)$ which boost the state in ${\hat n}$-direction by a speed $v$. Namely, the orthogonality is satisfied by the rotation, $R_{\hat n}^TR_{\hat n}=I$, but not by the Galilean boost, $G_{\hat n}^T(v)G_{\hat n}(v) \neq I$.   
While the properties of the rotation and the Galilean boost are characteristically different regarding helicity invariance, they both influence the orientation entanglement on the quantum states. As discussed in the case of spin-1 states, the rotation of the rest frame polarization vectors given by Eq.(\ref{spin-rotation}) exhibit the orientation entanglement
manifested by $\epsilon(+) \to \epsilon(-)$, $\epsilon(-) \to \epsilon(+)$, and $\epsilon(0) \to -\epsilon(0)$, respectively, under the rotation by $180^0$ around the $\hat{y}$-axis. 
As the spin itself is invariant under the Galilean boost, the same orientation entanglement can be identified in the boosted polarization vectors given by Eq.(\ref{GalvecP}), noting that
$\epsilon_g^\mu(P,+) \to 
-\epsilon_g^\mu(-P,+)$, $\epsilon_g^\mu(P,-) \to 
-\epsilon_g^\mu(-P,-)$, and $\epsilon_g^\mu(P,0) \to -\epsilon_g^\mu(-P,0)$, respectively, 
under the same rotation by $180^0$ around the $\hat{y}$-axis. Here, the momentum $\textbf{P}= P^1 \hat{x}+P^3\hat{z}$ undergoes the rotation by $180^0$ around the $\hat{y}$-axis, namely
$\textbf{P} \to -\textbf{P}$, 
while the helicity $\lambda$ is invariant under the rotation as discussed earlier. The sign flip of $\epsilon_g(P,\lambda) \to -\epsilon_g(-P,\lambda)$ signifies the orientation entanglement intrinsic to the rotation of spin-1 quantum states as discussed regarding Eq.(\ref{spin-rotation}). Both spin and momentum rotate together to flip the sign of the polarization vector $\epsilon_g(P,\lambda)$ while the helicity $\lambda$ remains same under the rotation. Moreover, we note that the identical quantum entanglement can be realized by the Galilean boost, which makes the transformation of the reference frame resulting $\textbf{P} \to -\textbf{P}$. It is manifest to see $\epsilon_g^\mu(-P,+) = \epsilon_g^\mu(P,-)$, $\epsilon_g^\mu(-P,-) = \epsilon_g^\mu(P,+)$, and $\epsilon_g^\mu(-P,0) = -\epsilon_g^\mu(P,0)$. It exhibits the non-invariance of the helicity under the Galilean boost, in contrast to the invariance of the helicity under the rotation.   

On the other hand, we know that Galilean boost should be modified to the Lorentz boost to satisfy relativistic causality. In particular,  two subsequent Lorentz boosts in different directions yield a rotation as exemplified by  Thomas precession~\cite{Thomas1926} in contrast to the Galilean boosts. Thus, one needs to reformulate the non-relativistic ``helicity" states into the relativistic ones by replacing the Galilean boosts by the Lorentz boosts. 
Such a real sense of helicity states that fulfill the relativistic causality were
described by Jacob and Wick \cite{JACOB1959404} with the sequence of operations discussed above, with the boost in the $\hat z$-direction followed by the rotation around the $\hat y$-axis, namely 
\begin{equation}	  
 	\label{RelT}
 T_I=  e^{-i \theta J^2 }e^{-i\eta {K}^3},
 \end{equation}
where $\theta$ is the angle parameter of the rotation around $\hat y$-axis and $\eta$ is the rapidity parameter of the boost in $\hat z$ direction. 
Following the sequence of operations on the rest frame spin polarization given by Eq.(\ref{PolaRest}) and taking the parameters $\theta = \tan^{-1}{(P^1/P^3)}$ and $\eta = \tanh^{-1}{(P^3/P^0)}$, one finds the so-called Jacob-Wick helicity states represented by the full relativistic polarization four vectors as given by
\begin{equation}
\label{LorvecP}
\begin{split}
\epsilon^{\mu}_I(P,+)= &- \Big(0, \frac{P^3}{|\textbf{P}|\sqrt{2}},   \frac{ i}{\sqrt{2}},\frac{ - P^1}{|\textbf{P}|\sqrt{2}}\Big),\\\\
\epsilon^{\mu}_I(P,-)= & \Big(0, \frac{P^3}{|\textbf{P}|\sqrt{2}},   -\frac{ i}{\sqrt{2}},\frac{ -P^1}{|\textbf{P}|\sqrt{2}}\Big),\\\\
\epsilon^{\mu}_I(P,0)=&  
\frac{1}{\sqrt{(P^0)^2-\textbf{P}^2}}
\Big(P^3, \frac{P^1 P^0}{|\textbf{P}|} ,0, \frac{P^3  P^0}{|\textbf{P}|} \Big),
\end{split}
\end{equation}
where $|\textbf{P}|=\sqrt{(P^1)^2+(P^3)^2}$ and the factor $\sqrt{(P^0)^2-\textbf{P}^2}$ should be kept for all regions of kinematics whether the particle is in the spacelike ($(P^0)^2<\textbf{P}^2$) region or timelike ($(P^0)^2>\textbf{P}^2$) region for the consistent normalization $\epsilon_I^\mu(P,\lambda)\epsilon_{I\mu}^*(P,\lambda)=-1$ for all $\lambda = \pm,0$. Here, we note that the Lorentz boost mixes the time and space components of a four-vector to yield the longitudinal polarization $\epsilon_I^\mu(P,0)$ in Eq.(\ref{LorvecP}) quite different from the non-relativistic longitudinal polarization $\epsilon_g^\mu(P,0)$ in Eq.(\ref{GalvecP}). Moreover, 
one may take $\sqrt{(P^0)^2-\textbf{P}^2}=M$ for the on-mass-shell particle and note the absence of the longitudinal polarization ($\lambda = 0$) for the massless ($M=0$) particle. 
This makes a remarkable difference in the relativistic Jacopb-Wick helicity satisfying the relativistic causality in contrast to the non-relativistic helicity discussed in Eq.(\ref{GalvecP}). 

 To obtain the helicity states, we first boost the spin in the $\hat{z}$-direction as shown in Eq. ($\ref{RelT}$). Thus, the longitudinal polarization vector changes its structure in the relativistic case, in contrast to the non-relativistic case, as discussed above. This change reflects the consistency of the relativistic quantum field theoretic formulation of the spin one particles with the spin one gauge field $A^\mu(p,\lambda)$ where $p$ and $\lambda$ represent the four-momentum and the helicity of the gauge field. In the radiation gauge, $A^0(p,\pm) = 0$ reflects the transverse polarization of the massless spin one particle, e.g., photon, which cannot possess the longitudinal polarization. Either the virtual photon with $p^2 \neq 0$ or the massive spin one particle, such as the $\rho$ meson ($p^2 = m_\rho^2$), acquires the longitudinal polarization with its time component dependent on the reference frame and non-zero if $|\textbf{P}| \neq 0$. Depending on the nature of the particle whether it's a spacelike or timelike virtual photon or a physical meson such as $\rho$, etc., one should take the mass $M$ in Eq.(\ref{LorvecP}) accordingly. 
In any case, the Jacob-Wick helicity polarization four vectors given by Eqs. (\ref{LorvecP}) are consistent with the polarization four vectors of the gauge fields in the radiation gauge $A^0=0$. 

Nevertheless, both the relativistic and non-relativistic helicities represented by $\epsilon_I^\mu(P,\lambda)$ and $\epsilon_g^\mu(P,\lambda)$ are invariant under the rotation. As exemplified for the non-relativistic helicity $\epsilon_g^\mu(P,\lambda)$ earlier under the rotation around $\hat{y}$-axis by $180^0$ , namely $\textbf{P} \to -\textbf{P}$,
the relativistic helicity undergoes
exactly the same transformation, 
$\epsilon_I^\mu(P,+) \to 
-\epsilon_I^\mu(-P,+)$, $\epsilon_I^\mu(P,-) \to 
-\epsilon_I^\mu(-P,-)$, and $\epsilon_I^\mu(P,0) \to -\epsilon_I^\mu(-P,0)$, exhibiting the identical sign flip signifying 
the orientation entanglement intrinsic to the rotation of spin-1 quantum states discussed earlier for $\epsilon_g(P,\lambda)$.    
As we noted for $\epsilon_g(P,\lambda)$, the identical quantum entanglement can be realized by the Lorentz boost as well
when it makes the transformation of the reference frame resulting $\textbf{P} \to -\textbf{P}$. It is manifest to see $\epsilon_I^\mu(-P,+) = \epsilon_I^\mu(P,-)$, $\epsilon_I^\mu(-P,-) = \epsilon_I^\mu(P,+)$, and $\epsilon_I^\mu(-P,0) = -\epsilon_I^\mu(P,0)$. It again exhibits the non-invariance of the helicity under the Lorentz boost, in contrast to the invariance of the helicity under the rotation. In this respect, it appears no difference in the characteristics of helicity between $\epsilon_I^\mu(P,\lambda)$ and $\epsilon_g^\mu(P,\lambda)$ with respect to the rotation and the boost, as well as the quantum entanglement aspect.   

However, the relativistic helicity given by Eq.(\ref{LorvecP}) is distinct from the non-relativistic helicity given by Eq.(\ref{GalvecP}), reflecting the difference due to the relativistic causality. 
In particular, the relativistic helicity zero polarization vector ($\epsilon_I^\mu(P,0)$) is drastically different from the non-relativistic one  ($\epsilon_g^\mu(P,0)$)
for the case of the fast moving particle helicity states.
Although the relativistic helicity $\pm$ polarization vectors ($\epsilon_I^\mu(P,\pm)$) look same as the non-relativistic ones 
($\epsilon_g^\mu(P,\pm)$),
in reality, they are also characteristically different due to the relativistic causality. 
While the characteristic of the non-relativistic helicity under the Galilean boost is immune to the mass of the particle, it's not the case for the relativistic helicity under the Lorentz boost.  
In the case of relativistic helicity, as the massless particle moves at the speed of light, its helicity cannot be flipped but remain same as nothing can move faster than the speed of light according to the relativistic causality. 
As discussed earlier in relation to the gauge field $A^\mu(p,\lambda)$,
the massless particle cannot possess the longitudinal polarization either. We may note that this property of helicity invariance for the massless particle leads ultimately to the chiral symmetry, as the helicity of a massless particle is the same as its chirality. 

As discussed earlier
in the back-to-back decay modes of the initial scalar state into two final spin-1 particles, the relative sign difference between 
$|1,\pm>|1,\mp>$ states and $|1,0>|1,0>$ state reveals the quantum orientation entanglement properties both in non-relativistic Galilean helicity formulation and relativistic Jacob-Wick helicity formulation. 
However, the nature of orientation entanglement in spin-1 systems may differ between the two formulations. Although the sign difference between $|1,\pm>|1,\mp>$ states and $|1,0>|1,0>$ state appears same in both formulation, the magnitude of decay rate to the $|1,0>|1,0>$, namely, helicity zero-zero state differs as the initial energy, equivalently the energy of the final spin-1 helicity zero particles become large. Such difference between the Galilean vs. Jacob-Wick helicity formulations can be detailed in the 
analysis of the relativistic scattering helicity amplitudes of $SS \to VV$, namely the two vector part production process in the annihilation of the two scalar particles.

Further difference between the Galilean helicity formulation and the relativistic Jacob-Wick helicity formulation can be found from the Wigner rotation~\cite{Wigner1939} effect in the transverse boost which appears in the Lorentz transformation but not in the Galilean transformation. Such difference would affect the physical appearance of the orientation entanglement effect in details. As the Wigner rotation effect depends on the transverse boost parameters, it is rather complicate to analyze the details of such effect. For this reason, the Jacob-Wick helicity was formulated from the first place without using any transverse boost, but the longitudinal boost with the subsequent rotation as shown in Eq.(\ref{RelT}) to preserve the identity of the helicity. Such observation motivates us to look for an alternative relativistic helicity formulation which takes care of the complicate Wigner rotation effect automatically such that it preserves its own identity of the helicity whichever direction of the boost is applied. As we discussed in the introduction, there exists different forms of relativistic dynamics~\cite{RevModPhys.21.392} and in particular the LFD provides the maximum number (seven) of kinematic generators in the Poincar\'e group. It turns out that the advantage of the maximum number of kinematic generators in LFD achieves the goal of finding the alternative helicity formulation preserving its own identity of the helicity whichever direction of the boost is applied. 

To define the light-front (LF) helicity states, we use the following transformation;
	\begin{equation}	  
 	\label{LFT}
 T_{l}= e^{-i \beta_1  E^1}e^{-i\beta_3 {K}^3},
 \end{equation}
 where $E^1 = (J^2+ K^1) / \sqrt{2}$ is the LF transverse boost and $K^3$ is the longitudinal boost. Both of these operators are kinematic operators at the light front, in the sense that the $x^+=0$ plane is preserved under the transformations generated by them. 
In particular, one should note that the LF transverse boost $E^1 = (J^2+ K^1) / \sqrt{2}$ involves both 
 the transverse rotation and boost together to preserve the $x^+=0$ plane invariant. The entire transformation given by Eq.(\ref{LFT}) is thus kinematic and provides the invariance of the helicity regardless of the boost directions. Effectively, the light-front helicity formulation takes care of the complicate Wigner rotation effect discussed in the Jacob-Wick helicity formulation earlier. 

 To derive LF polarization vectors starting from the polarization vectors given by Eq.(\ref{PolaRest}) in the rest frame,
we first apply the longitudinal boost $e^{-i\beta_3 {K}^3}$ in Eq.(\ref{LFT}) with the parameter $\beta_3 =\tanh^{-1} (P^3/P^0)$ which is identical to the rapidity parameter $\eta =\tanh^{-1} (P^3/P^0)$ given by Eq.(\ref{RelT}) in IFD. Using the LF longitudinal momentum $P^+ =( P^0+ P^3)/ \sqrt{2}$, this longitudinal rapidity parameter $\beta_3 = \eta$ can be denoted in terms of $P^+$ as 
$\beta_3 = \tanh^{-1} \big((\sqrt{2}P^+ -M)/(\sqrt{2}P^+ +M)\big)$.   
Then, applying the subsequent LF transverse boost $e^{-i \beta_1  E^1}$ in Eq.(\ref{LFT})
with the transverse LF transverse boost parameter 
$\beta_1 = P^1/ P^+$ provided by the ratio of $P^1$ with respect to $P^+$, we find the LF polarization four-vectors given by 
\begin{equation}
\label{LFPola1P}
\begin{split}
\epsilon^{\mu}_L(P,+)= &- \Big(\frac{P^1}{2P^+ }, \frac{1}{\sqrt{2}},   \frac{ i}{\sqrt{2}},\frac{-P^1}{2P^+ }\Big),\\
\epsilon^{\mu}_L(P,-)= & \Big(\frac{P^1}{2P^+ }, \frac{1}{\sqrt{2}},   -\frac{ i}{\sqrt{2}},\frac{-P^1}{2P^+ }\Big),\\
\epsilon^{\mu}_L(P,0)=& \Big(\frac{2(P^+)^2- M^2+(P^1)^2}{2\sqrt{2}P^+ M},   \frac{P^1}{M},   0,\\&\frac{2(P^+)^2+ M^2-(P^1)^2}{2\sqrt{2}P^+ M}\Big),\\
\end{split}
\end{equation}

where the four-vectors are denoted in 
the basis of ($\epsilon ^0 , \epsilon^1 , \epsilon^2 , \epsilon^3 $) with $\mu = 0,1,2,3$. To derive the LF variables for the polarization four-vectors in the basis of ($\epsilon^+ , \epsilon^1 , \epsilon^2 , \epsilon^- $), we may consider $(G^{\hat{\nu}}_{\mu})_{\delta \rightarrow \pi/4}$ in Eq. (\ref{InterCorU}).Thus the LF space-time transformation matrix can be written as,
   \begin{equation}
    \label{LFspT}
G^{\nu}_{\mu}
	=
	\begin{pmatrix}
	\frac{1}{\sqrt{2}} & 0 & 0 & \frac{1}{\sqrt{2}}\\
	0 & 1 & 0 & 0 \\
	0 & 0 & 1 & 0 \\
	\frac{1}{\sqrt{2}} & 0 & 0 & -\frac{1}{\sqrt{2}}	\\
	\end{pmatrix}.
	\end{equation} 
After applying $G^{\nu}_{\mu}$  matrix on polarization four-vectors in the basis of ($\epsilon^0 , \epsilon^1 , \epsilon^2 , \epsilon^3 $) in Eq. (\ref{LFPola1P}) we can get the LF polarization for vectors $\epsilon^{\nu}_L (P,\lambda) =G^{\nu}_{\mu} \epsilon^{\mu}_L (P,\lambda) $ in the basis of $\nu = +,1,2,- .$   
  \begin{equation}
\label{LFPola2}
\begin{split}
\epsilon^{\nu}_L(P,+)= &- \Big(0, \frac{1}{\sqrt{2}},   \frac{ i}{\sqrt{2}},\frac{  P^1}{\sqrt{2} P^+}\Big),\\
\epsilon^{\nu}_L(P,-)= & \Big(0, \frac{1}{\sqrt{2}},   -\frac{ i}{\sqrt{2}},\frac{ P^1}{\sqrt{2} P^+}\Big),\\
\epsilon^{\nu}_L(P,0)=& \Big(\frac{P^+}{M},   \frac{ P^1}{M},   0,\frac{(P^1)^2-M^2}{2P^+ M}\Big).\\
\end{split}
\end{equation}

Having discussed both the Jacob-Wick helicity polarization four-vectors given by Eq.(\ref{LorvecP}) and the LF helicity polarization four-vectors given by Eq.(\ref{LFPola2}), we now explore the correspondence between them
with the interpolation between IFD and LFD, parametrized by the interpolation angle, $\delta$, which was introduced in Eq,(\ref{InterCorU}). As the detailed derivation of the interpolating helicity polarization four-vectors that link between Eq.(\ref{LorvecP}) and Eq.(\ref{LFPola2}) has already been presented in our earlier work~\cite{PhysRevD.91.065020}, we first briefly summarize our derivation in the next section, Sec.\ref{Sec:IV}, 
and discuss the physical spin orientation relative to the momentum in the interpolating helicity eigenstates to analyze the orientation entanglement behaviors with respect to the interpolation angle $\delta$. 
With the interpolating helicity formulation between IFD and LFD, we will discuss the quantum orientation entanglement properties in the relativistic scattering helcity amplitudes of $SS \to VV$ for the entire region ($0 \le \delta \le \pi/4$) of the dynamics between IFD ($\delta =0$) and LFD ($\delta = \pi/4)$. \\

\section{ Spin Orientation of the Interpolating Helicity States}
	\label{Sec:IV}

To incorporate the interpolation angle, we utilize the kinematic generators $\mathcal{K}^{\hat{1}}$ and $\mathcal{K}^{\hat{2}}$ given by $\mathcal{K}^{\hat{1}} = -K^1 \sin\delta -J^2 \cos\delta$ and $\mathcal{K}^{\hat{2}} = J^1 \cos\delta -K^2 \sin\delta$ which coincide with transverse rotation ( $J^1$ , $J^2$ ) in IFD ($\delta \rightarrow 0$) and with transverse boost ($E^1$ , $E^2 $ ) in LFD ($\delta \rightarrow \pi/4$) ~\cite{PhysRevD.91.065020} The generalized interpolation transformation matrix is given by ~\cite{PhysRevD.91.065020}, 
	\begin{equation}
		\label{Tdelta1}
	T_{\delta}= e^{i\beta_1 \mathcal{K}^{\hat{1}} + i\beta_2 \mathcal{K}^{\hat{2}} } e^{-i\beta_3 {K}^3}, 
	\end{equation}
where all three parameters  $\beta_1,\beta_2$ and $\beta_3$ are kept for generality. 
Note here that Eq.(\ref{RelT}) 
in IFD with $\beta_1 = \theta $, $\beta_2=0$ and Eq.(\ref{LFT}) in LFD with $\beta_2=0$. 

The values $\beta_1 ,\beta_2, \beta_3 $ for an interpolation angle $\delta$ are related to the desired final momentum $\vec{P} =(P^1 , P^2, P^3)$ of the particle with mass M . The detailed derivation of the relationship between $(\beta_1, \beta_2, \beta_3)$ for a given  $ \delta$ and the 4-momentum component $P^{\hat{\mu}}$ has been worked out in our previous work ~\cite{PhysRevD.91.065020} and summarized in Appendix~\ref{App: A}.

After applying the interpolating $T _{\delta}$ transformation  to the rest frame helicity states in Eq. (\ref{PolaRest}) , we can derive the interpolating polarization vectors $\epsilon^{\mu}_{\delta}(P,\lambda)$;

\onecolumngrid
\begin{equation}
	\label{Pola delta1}
	\begin{split}
	\epsilon^{\mu}_{\delta}(P,+)=&-\frac{1}{\sqrt{2}\mathbb{P}}\Big(\sin\delta\, | \textbf{P}_{\perp}|, \frac{(P^1 P_{\hat{-}}-{\it{i}} P^2 \mathbb{P})}{| \textbf{P}_{\perp}|},\frac{(P^2 P_{\hat{-}}+{\it{i}} P^1 \mathbb{P})}{| \textbf{P}_{\perp}|}, -\cos\delta\, {| \textbf{P}_{\perp}|}\Big),\\
	\epsilon^{\mu}_{\delta}(P,-)= &\frac{1}{\sqrt{2}\mathbb{P}}\Big(\sin\delta\, | \textbf{P}_{\perp}|, \frac{(P^1 P_{\hat{-}}+{\it{i}} P^2 \mathbb{P})}{| \textbf{P}_{\perp}|},\frac{(P^2 P_{\hat{-}}-{\it{i}} P^1 \mathbb{P})}{| \textbf{P}_{\perp}|},-\cos\delta\, {| \textbf{P}_{\perp}|}\Big),\\
	\epsilon^{\mu}_{\delta}(P,0)=& \frac{1}{M\mathbb{P}}\Big(\frac{\mathbb{P}^2 \cos\delta -P_{\hat{-}}P^{\hat{+}} \sin\delta}{\mathbb{C}} ,P^1 P^{\hat{+}}, P^2 P^{\hat{+}}, \frac{P_{\hat{-}}P^{\hat{+}} \cos\delta -\mathbb{P}^2 \sin\delta}{\mathbb{C}} \Big),\\
	\end{split}
	\end{equation} 
\twocolumngrid  	
\noindent
where $\mathbb{P}=\sqrt{(P^{\hat{+}})^2-M^2\mathbb{C}}=\sqrt{(P_{\hat{-}})^2+ | \textbf{P}_{\perp}|^2 \mathbb{C} } $ and the four-vectors are  denoted in the basis of $(\epsilon ^0 , \epsilon^1 , \epsilon^2 , \epsilon^3) $ with   $\mu = 0,1,2,3 $. To recover IFD polarization vectors given in Eq. (\ref{LorvecP}), we can take the limit of  $\delta \rightarrow 0$  to get  $P^{\hat{+}} \rightarrow P^0 $ , $P_{\hat{-}} \rightarrow  P^3$   $\mathbb{P} \rightarrow  |P|$  and  $\mathbb{C} \rightarrow  1$ with   $P^2=  0$ as $\beta_2 =0$.  Similarly in the LFD limit we can recover the LF polarization vectors given in Eq. (\ref{LFPola1P} ) in the basis of $(\epsilon ^0 , \epsilon^1 , \epsilon^2 , \epsilon^3) $ with $P^2 =0$. Since in the  LFD ($\delta \rightarrow \pi/4$), $\mathbb{C} \rightarrow 0$ and $\mathbb{P} \rightarrow P ^+ $, $P ^{\hat{+}} \rightarrow P ^+ $,$ P_{\hat{-}} \rightarrow P ^+  $ make the denominator as well as the numerator go to zero to get the finite  $\epsilon ^0 $ and $\epsilon ^3 $ components of  $\epsilon^{\mu}_{\delta=\pi/4}(P,0)$  which can be seen in Eq. (\ref{LFPola1P}).

We use the relation $x^{\hat{\mu}}$ = $G^{\hat{\mu}}$$_\nu $$x^\nu $ given in Eq. (\ref{InterCorU}) and the polarization vectors $\epsilon^{\mu}_{\delta}(P,\lambda)$ given in Eq. (\ref{Pola delta1}) to find the interpolating basis component of interpolating polarization vectors $\epsilon^{\hat{\mu}}(P,\lambda)= G^{\hat{\mu}} _{\nu }\epsilon^\nu_{\delta} (P,\lambda) = (\epsilon^{\hat{+}} , \epsilon^1 , \epsilon^2 , \epsilon^{\hat{-}} )$ given below, 

\onecolumngrid

 \begin{equation}
	\label{Pola delta2}
	\begin{split}
	\epsilon^{\hat{\mu}}(P,+)=&-\frac{1}{\sqrt{2}\mathbb{P}}\Big(0 , \frac{(P^1 P_{\hat{-}}-{\it{i}} P^2 \mathbb{P})}{| \textbf{P}_{\perp}|},\frac{(P^2 P_{\hat{-}}+{\it{i}} P^1 \mathbb{P})}{| \textbf{P}_{\perp}|}, {| \textbf{P}_{\perp}|}\Big),\\
	\epsilon^{\hat{\mu}}(P,-)=&\frac{1}{\sqrt{2}\mathbb{P}}\Big(0, \frac{(P^1 P_{\hat{-}}+{\it{i}} P^2 \mathbb{P})}{| \textbf{P}_{\perp}|},\frac{(P^2 P_{\hat{-}}-{\it{i}} P^1 \mathbb{P})}{| \textbf{P}_{\perp}|}, {| \textbf{P}_{\perp}|}\Big),\\
	\epsilon^{\hat{\mu}}(P,0)= &\frac{1}{M\mathbb{P}}\Big( \mathbb{P}^2 ,P^1 P^{\hat{+}}, P^2 P^{\hat{+}}, P^{\hat{-}}P^{\hat{+}}  -M^2 \mathbb{S} \Big).
	\end{split}
	\end{equation}

\twocolumngrid

In the IFD limit ($\delta \rightarrow 0$), Eq. (\ref{Pola delta2}) will coincide with the IFD polarization vectors given in Eq. (\ref{LorvecP}), while in the LF limit ($\delta \rightarrow \pi /4 $), they  coincide with Eq. (\ref{LFPola2}) where we have light-front polarization vectors in the basis of ($\epsilon^+ , \epsilon^1 , \epsilon^2 , \epsilon^- $). Note that this interpolating polarization vectors $\epsilon^{\hat{\mu}}(P,\lambda)$ respects the  gauge condition $A^{\hat{+}}=0$ and $\partial_{\hat{-}}+ \partial_\perp \cdot A_\perp\mathbb{C}=0$ which links Coulomb gauge $A^0=0$ and $\nabla \cdot A=0$ in IFD and the light-front gauge $A^+=0$ in the LFD as discussed in Ref ~\cite{PhysRevD.91.065020}. Besides that the interpolating polarization vectors satisfy the transversality and orthogonality constraints ($\epsilon_{\hat{\mu}}(P,\lambda)P^{\hat{\mu}}=0 $ and $ \epsilon^*(P,\lambda) .\epsilon(P,\lambda')=-\delta_{\lambda,\lambda'} $ ) as expected.

To confirm the helicity eigenvalues of each interpolating , spin-1 polarization vector one may use the generalized interpolating operator \cite{PhysRevD.92.105014} given by ;
	
		\begin{equation}
	 \mathcal{J}^3_{\delta} = T_{\delta}J^3 T^{-1}_{\delta} = J^3\cos \alpha + (\beta_1 \mathcal{K}^{\hat{2}} - \beta_2 \mathcal{K}^{\hat{1}})\frac{\sin\alpha}{\alpha},
		\end{equation}
with  $\alpha= \sqrt{\mathbb{C} (\beta_1 ^2 +\beta_2 ^2)}$.  After applying the corresponding interpolating four vector momenta, we can write it as given by
	\begin{equation}
\label{J3delta}
	 \mathcal{J}^3_{\delta} = \frac{1}{\mathbb{P}}\big[P_{\hat{-}} J^3 +P^1 \mathcal{K}^{\hat{1}} -P^2 \mathcal{K}^{\hat{2}} \big]. 
	\end{equation}
In the instant form limit ($ \delta \rightarrow 0$), $\mathcal{K}^{\hat{1}} \rightarrow -J^2 ,\mathcal{K}^{\hat{2}} \rightarrow J^1 $ , $P_{\hat{-}} \rightarrow P^3 $ and $ \mathbb{P} \rightarrow \sqrt{ (P^0)^2 -M^2} =|\bf{P}| , $  the operator $\mathcal{J}^3_{\delta}$ coincide with the familiar  Jacob and Wick operator  $\bf{P.J}/ |\bf{P}|$ . In the light-front limit ($ \delta \rightarrow \pi/4$),  $ K^{\hat{1}} \rightarrow -E^1 , K^{\hat{2}} \rightarrow -E^2 ,P_{\hat{-}} \rightarrow P^+ $ and $ \mathbb{P} \rightarrow  P^+ $,  and thus the operator coincide with light-front helicity operator $ J^3 + (P^2E^1 -P^1 E^2)/P^+ $ as we expected.

Then, when we apply any eigenvector state, such as interpolating polarization vectors in the basis of $(\epsilon ^0 , \epsilon^1 , \epsilon^2 , \epsilon^3) $ such that, $\mathcal{J}^3_{\delta} \epsilon^{\mu}_{\delta}(P,\lambda) = \lambda \epsilon^{\mu}_{\delta}(P,\lambda),$ we can recover the interpolating helicity eigenvalue (  $\lambda $)  of the corresponding helicity state. The same equation can be written in the interpolating basis,

		\begin{equation}
        \hat{\mathcal{J}}_3
         \epsilon^{\hat{\mu}}(P,\lambda) = \lambda \epsilon^{\hat{\mu}}(P,\lambda),
	\end{equation}
where $ \hat{\mathcal{J}}_3 =(G^{\hat{\mu}} _{\nu }) \mathcal{J}^3_{\delta} (G^{\hat{\mu}} _{\nu })^{-1} $. One may note that the helicity operator $\hat{\mathcal{J}}_3 $ can also be written in terms
of the Pauli-Lubanski operator~\cite{LEUTWYLER197894} ; $W^\mu = \frac{1}{2}\, \, \epsilon^{\mu\nu\alpha\beta}\, P_\nu\, M_{\alpha\beta}$ simply as $\hat{\mathcal{J}}_3=W^{\hat{+}}/\mathbb{P}$.
	
In the IFD ordinary notion of helicity is usually defined by the spin parallel or antiparallel to the particle momentum direction. For different interpolation angles, in general, there is a relative angle between the spin orientation and the momentum direction. In LFD, the transverse light-front boost
operators $E^1$ and $E^2$ involve the rotations, and they generate the angle between the spin orientation and the momentum direction. In the next subsection, we will discuss this difference between spin and momentum direction depending on the interpolation angle. 

\subsection{ Spin Orientation Relative to the Momentum Direction}
		\label{Sec:IV.A}
	
To find the spin orientation changes with the particle's momentum direction and the interpolation angle, we rewrite the $T_{\delta}$ transformatio in Eq. (\ref{Tdelta1}) transformation as follows.
	
	\begin{equation}
	\label{Tdelta2}
	T_\delta= B(\eta)D(\hat{m}, \theta_s)= e^{-i\eta K} e^{-i \hat{m}J \theta_s}
	\end{equation}

The $ D(\hat{m}) =e^{-i \hat{m}J \theta_s} $ operator rotates the spin around the axis of unit vector $\hat{m}=( -\sin\phi_s, \cos\phi_s ,0)$ by angle $ \theta_s$ and then the operator $ B(\eta)=e^{-i\eta K} $ boost the spinor to momentum $\bf{P} $ along the unit vector $\hat{n}=(\sin\theta \cos\phi , \sin\theta \sin\phi ,\cos\theta) $.
	
Comparing the explicit matrix form of $T_{\delta}$ transformation given by Eq.(\ref{Tdelta1}) and Eq.(\ref{Tdelta2}) , we find the relations between the spin angles $(\theta_s, \phi_s) $ and  the interpolating  momentum variables as below,

	\begin{equation}
	\label{thetas}
	\begin{split}	
	\cos\theta_s =& \frac{\cos\alpha (1+\cosh\beta_3)+\cosh\beta_3 - \cosh\eta}{1+ \cosh\eta}, \\
	=&\frac{P_{\hat{-}}}{\mathbb{P}} -\frac{\bf{P}_{\perp}^2 \sin\delta}{\mathbb{P} (E+M)},
	\end{split}
	\end{equation}
	\begin{equation}
	\label{phis}
	\begin{split}
	\cos\phi_s = & \frac{\beta_1^2}{\sqrt{\beta_1^2 +\beta_2^2}}= \cos\phi=\frac{P^1}{\sqrt{\bf{P}_{\perp}^2 }}, \\
	\sin\phi_s =& \frac{\beta_2^2}{\sqrt{\beta_1^2 +\beta_2^2}}= \sin\phi=\frac{P^2}{\sqrt{\bf{P}_{\perp}^2 }},
	\end{split}
	\end{equation}
where we use the interpolating dynamic relation given in the Appendix ~\ref{App: A} along with the  $\sinh(\eta) = |\bf{P}|$$ /M$ and $\cosh(\eta) = E/M$, to find the interpolating four momentum dependence. Here, $\phi_s=\phi$ as the particle's motion direction and the spin direction are in the same plane, taken as the $x-z$ plane, i.e $\phi_s=\phi=0$.

\begin{center}
	 	\begin{figure}[H]	 	\includegraphics[width=8cm]{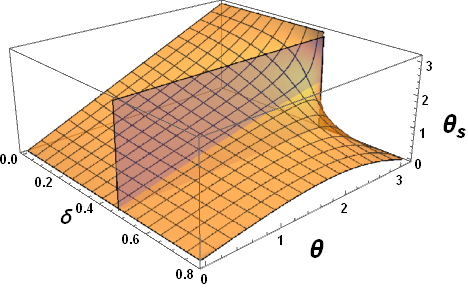}
		\caption{Spin orientation ($\theta_s$) dependence on the particle's momentum direction ($\theta$) and the interpolation angle ($\delta$) for the helicity plus state }
			\label{Fthetas}
		\end{figure}
	\end{center}

If we consider a vector particle with four momenta $P^{\mu}= ( E_0, P_v \sin\theta, 0, Pv \cos\theta )$, using Eq.(\ref{thetas}), we can write the $\cos\theta_s$ explicitly for a helicity plus state as 

\begin{equation}
\label{thetas1}
    \begin{split}
    &\cos\theta_s=\\
        &\frac{
P_v \cos\delta \cos\theta
+ \sin\delta \left(
E_0 - \left(E_0 - \sqrt{E_0^2 - P_v^2}\right)\sin^2\theta
\right)
}{
\sqrt{
\left(P_v \cos\delta \cos\theta + E_0 \sin\delta\right)^2
+ P_v^2 \mathbb{C} \sin^2\theta
}
},
    \end{split}
\end{equation}
where $E_0 = \sqrt{P_v^2+M_v^2}$ with  $P_v$ and  $M_v$ that are the momentum and mass of the vector particle, respectively. In Fig~\ref{Fthetas}, we display $\theta_s$ in terms of $\theta$ and $\delta$. Here, we took the energy $(E_0)$ and the momentum $(P_v)$ values of the particle as $2$ GeV and $1$ GeV, respectively.

    \vspace{-1cm}
As we consider the helicity plus state, we can see that the spin orientation and the momentum direction are parallel ($\theta =\theta_s$) to each other, which yields a straight line in Fig.  \ref{Fthetas} in the IFD that satisfies the Jacob and Wick helicity definition. That also can be calculated using Eq. (\ref{thetas1}) as $\delta \rightarrow 0 $ , $\cos\theta_s = \cos\theta$.

Besides that, we can see that for any interpolation angle when $\theta \rightarrow 0$, $\theta_s \rightarrow 0$, meaning that there is no difference between spin and momentum directions when particle is moving the $+z$ direction.

When we increase the interpolation angle up to some specific value of ($\delta_c$), we see spin direction ( $ \theta_s $ ) following the momentum direction ( $ \theta $ ), similar to the IFD helicity. But after that value ($\delta_c <\delta \leq \pi/4$), $\theta_s$ does not follow $\theta$ all the way to $\pi$, but instead starts to decrease beyond a certain point goes back to 0 when $\theta= \pi $. One can calculate this  $\delta_c$ value by considering $\theta= \pi$, in Eq. (\ref{thetas1}) and, $\bf{P}_{\perp} \rightarrow 0$ in Eq. (\ref{thetas})

\begin{equation}
\label{thetas1Pi}
    (\cos\theta_s)_{\theta \rightarrow \pi}= \frac{P_{\hat{-}}}{|P_{\hat{-}}|} =
        \frac{
 E_0 \sin\delta - P_v \cos\delta  
}{
\sqrt{
E_0 \sin\delta - P_v \cos\delta  }},
\end{equation}

\begin{center}
	 	\begin{figure}[H]	 	\includegraphics[width=8cm]{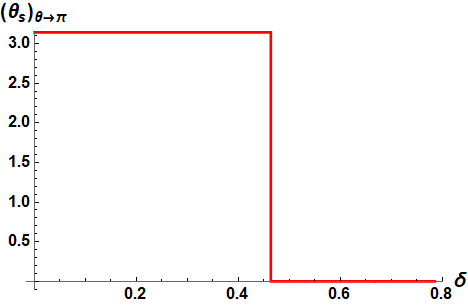}
		\caption{Spin direction dependence on the interpolation angle of helicity plus state when the particle is moving in the $-z$ direction.}
			\label{thetaspi}
		\end{figure}
	\end{center}

\begin{center}
	 	\begin{figure}[H] 	\includegraphics[width=8cm]{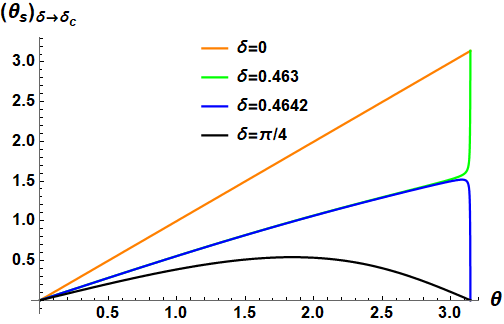}
		\caption{Spin direction ($\theta_s$) dependence on particle moving direction ($\theta$) just before and after $\delta_c$, along with $\delta=0$ and $\delta=\pi/4$ profiles.}  	\label{thetasdeltac1}
		\end{figure}
	\end{center}

According to Eq. (\ref{thetas1Pi}), the sign change in $P_{\hat{-}}=P_v \cos\delta 
- E_0 \sin\delta $, corresponds to the critical interpolation angle that can be found as  $ \delta_c = \tan^{-1} (P_v/E_0)$ at the condition of $P_{\hat{-}}=0$. The $\delta_c$ value can be seen in the figure when $\delta =\tan^{-1}(1/2)$. Fig. \ref{thetaspi} shows the profile of $\theta_s$ when $\theta \rightarrow \pi$ where the spin direction abruptly changes $180^0$ before and after the critical interpolation angle $\delta_c =\tan ^{-1} (1/2) \approx 0.4636$.
\vspace{-1cm}
To examine the spin direction dependence on the particle moving direction just before ($\delta=\delta_c-\epsilon$) and after ($\delta=\delta_c+\epsilon$) the interpolation angle $\delta_c \approx 0.4636$, we exhibit the profiles of the spin direction for $\delta=0.463$ and $\delta=0.4642$ taking $\epsilon \approx 0.0006$ in Fig~\ref{thetasdeltac1}. A dramatic profile change can be seen just before and after $\delta_c$ compared with the IFD ($\delta =0)$ and LFD ($\delta=\pi/4$) profiles. 
\vspace{-1cm}
 Therefore, this specific value of interpolation angle ($\delta_c$) clearly bifurcates the branch that belongs to LFD from the branch that belongs to IFD. That bifurcation can be seen as the transparent vertical plane in Fig.~\ref{Fthetas} consecutively in the profile of $(\theta_s)_{\theta \rightarrow \pi}$  in Fig.~\ref{thetaspi} as well as in the Fig.~\ref{thetasdeltac1}. The more details related to this bifurcation have been discussed in the previous work~\cite{PhysRevD.92.105014}.

		\subsection{ Interpolating Wigner-d matrix of helicity states}
		\label{Sec:IV.B}
				
As we are interested in seeing the orientation entanglement of helicity states, and we realized that there are relative angle differences between the spin direction and the momentum, which depend on the interpolation angle, we want to see how that difference manifests within the interpolating helicity states between IFD and LFD. Reference helicity states can be chosen as the Jacob and Wick helicity in the IFD, as the relative difference between spin and momentum directions is well defined. As an example, for the Jacob-Wick plus/minus helicity state, the momentum direction is always parallel/anti-parallel to the spin direction, while for the zero-helicity state, they are perpendicular to each other.   

Analyzing interpolating spin-1  polarization vectors and their IFD helicity states, we found that the interpolating helicity states can be expanded in the basis of their IFD helicity states using the Wigner-d type of function that depends on the interpolation angle. The relations can be summarized as below,

\begin{widetext}
	\begin{equation}
\label{HWM 1 Pola}
\begin{split}
\epsilon^{\mu}_{\delta}(P,+)=&(\frac{1+\cos\theta_h }{2})\epsilon^{\mu}_{I} (P,+)+ \frac{\sin\theta_h }{\sqrt{2}}\epsilon^{\mu}_{I} (P,0)+(\frac{1-\cos\theta_h }{2})\epsilon^{\mu}_{I} (P,-),\\
\epsilon^{\mu}_{\delta}(P,0)=& - \frac{\sin\theta_h }{\sqrt{2}}\epsilon^{\mu}_{I} (P,+)+ \cos\theta_h\epsilon^{\mu}_{I} (P,0)+\frac{\sin\theta_h }{\sqrt{2}} \epsilon^{\mu}_{I} (P,-), \\
\epsilon^{\mu}_{\delta}(P,-)=&(\frac{1-\cos\theta_h }{2})\epsilon^{\mu}_{I} (P,+)- \frac{\sin\theta_h }{\sqrt{2}}\epsilon^{\mu}_{I} (P,0)+(\frac{1+\cos\theta_h }{2})\epsilon^{\mu}_{I} (P,-),
\end{split}
\end{equation}
\end{widetext}
where $ \theta_h= (\theta_s - \theta) $ and $\epsilon^{\mu}_{\delta} (P,\lambda) = G^{\mu} _{\hat{\nu} } \epsilon^{\hat{\nu}} (P,\lambda)$ . We can obtain $\theta_s $ from Eq. (\ref{thetas}), which represents the spin direction for a corresponding momentum direction and interpolation angle. We may identify $\theta_h$ as the interpolating helicity angle, which directly represents the relative angle between spin and momentum. Similar to Eq. (\ref{thetas}), we can write the $\theta_h$ relation in terms of interpolating four momentum as
\begin{equation}
    \theta_h =-\cos^{-1} \Big[ \frac{\mathbb{P}^2 \cos\delta -P ^{\hat{+}}P_{\hat{-}}\sin\delta}{\mathbb{P} \mathbb{C}|\textbf{P}|}\Big].
\end{equation}
 In the IFD ( $\delta \rightarrow 0$) , as  $\mathbb{P} \rightarrow  |P|$  and  $\mathbb{C} \rightarrow  1$ , $\theta_h =0$. But in the LFD ($\delta \rightarrow \pi/4$), $\mathbb{C} \rightarrow 0$ and $\mathbb{P} \rightarrow P ^+ $, $P ^{\hat{+}} \rightarrow P ^+ $,$ P_{\hat{-}} \rightarrow P ^+  $ make the denominator as well as the numerator go to zero in 
 
\begin{equation}
\label{LF finite}
\Big[\frac{\mathbb{P}^2 \cos\delta -P ^{\hat{+}}P_{\hat{-}}\sin\delta}{\mathbb{P} \mathbb{C}|\textbf{P}|}\Big] _{\delta \rightarrow \pi/4} =\frac{P^+(P^+ -P ^- )+ |P_{\perp}^2|}{\sqrt{2}P^+ |\textbf{P}|},
\end{equation}
which gives a finite value $\theta_h$ in the light front. 

\begin{center}
	 	\begin{figure}[H]	 	\includegraphics[width=8cm]{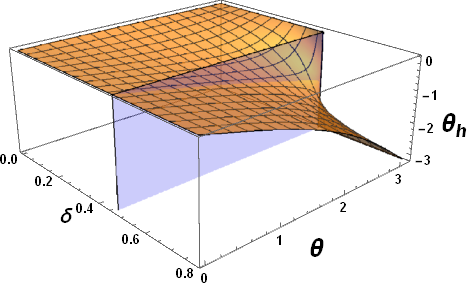}
	\caption{Interpolating helicity angle dependence on the particle's momentum direction for the plus helicity state.}
			\label{Fthetah}
		\end{figure}
	\end{center}

If we use the same  four vector momentum we use in the above subsection $P^{\mu}= ( E_0, P_v \sin\theta, 0, Pv \cos\theta )$, we can find out the corresponding $\theta_h$, 

\begin{equation} 
\label{thetaheq}
\theta_h=
-\cos^{-1} \Big[\frac{P_v \cos\delta + E_0 \cos\theta \sin\delta}
{\sqrt{\left(P_v \cos\delta \cos\theta + E_0 \sin\delta\right)^2
+ P_v^{\,2}\mathbb{C}\sin^2\theta}}\Big].
\end{equation}
Fig.~\ref{Fthetah} shows the interpolating helicity angles changes with the interpolation angle and the particle's momentum direction for the plus helicity state where we used relationship we derived in Eq. (\ref{thetaheq}). Here, we also took the energy $(E_0)$ and the momentum $(P_v)$ values of the particle as $2 $ GeV and $1$ GeV, respectively.

    \vspace{-1cm}
We can see that, in Eq. (\ref{thetaheq}) and Fig \ref{Fthetah},
In the the limit $\delta \rightarrow 0$ ,  $\theta_h \rightarrow 0$ as we expected in the plus helicity state, since $\theta_s= \theta$ in the IFD. In addition, for any interpolation angle when $\theta=0$, $(P_v \cos\delta + E_0 \sin\delta) >0 $ thus $\theta_h \rightarrow 0$.

However, when $\theta=\pi$, Eq.~(\ref{thetaheq}) becomes 
\begin{equation} 
\label{thetaheq1}
\cos(-\theta_h)=
 \frac{P_v \cos\delta - E_0  \sin\delta}
{|P_v \cos\delta - E_0  \sin\delta|} ,
\end{equation}
where we see the division depending on the sign of the ($ P_v \cos\delta - E_0  \sin\delta$). In the region of  $ 0<\delta < \delta_c$, $( E_0 \sin\delta - P_v \cos\delta ) > 0 $, $\theta_h \rightarrow 0$ while in the region of  $ \delta_c < \delta < \pi/4$, $\theta_h \rightarrow -\pi$. Therefore, Fig~ \ref{Fthetah}  also  show that the $\delta_c$ bifurcates the branch that belongs to LFD from the branch that belongs to IFD.
We have shown that division as a transparent vertical plane in Fig.~\ref{Fthetah} as well.

\begin{center}
		\begin{figure}[H]	\includegraphics[width=4cm]{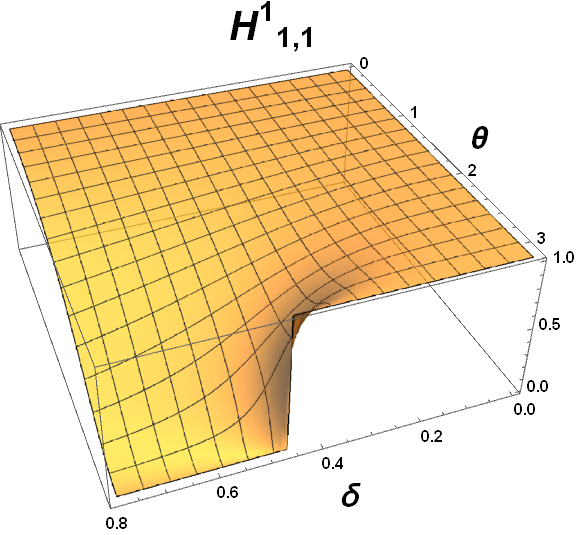}
			\includegraphics[width=4cm]{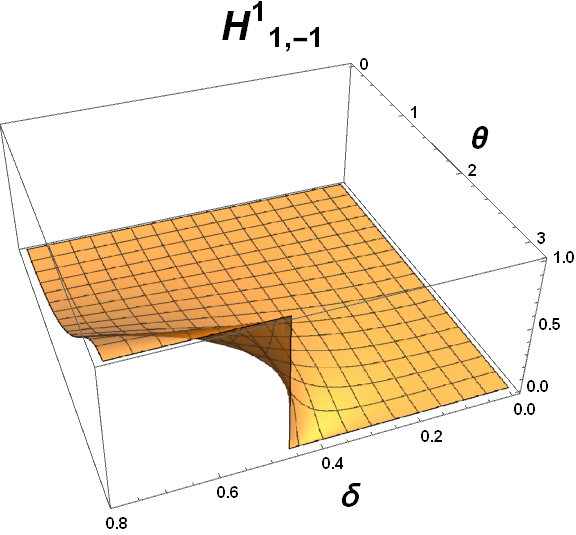}\\
			\includegraphics[width=4cm]{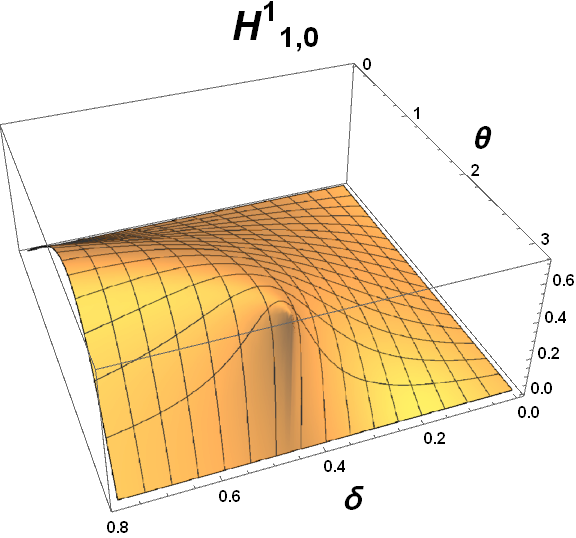}
			\includegraphics[width=4cm]{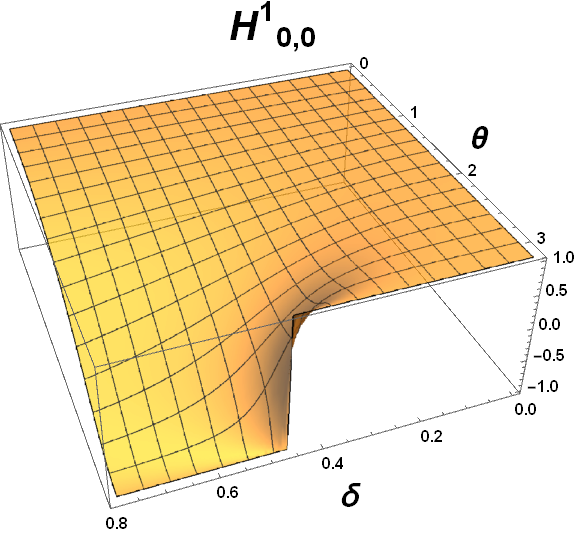}\\
			\caption{ Independent Wigner matrix elements $H^{1}_{\lambda,\lambda}$ of spin-1 helicity polarization vectors that depend on the interpolation angle and the particle's moving direction. Other Wigner matrix elements can be related by Eq.~(\ref{HWMP 1}). }
			\label{FHWM 1}
		\end{figure} 
	\end{center}

We may now summarize Eq.(\ref{HWM 1 Pola}) as 
\begin{equation}
\label{summary HWM 1 Pola}
\epsilon^{\mu}_{\delta} (P,\lambda)= \sum_{\lambda'=-1}^{1} H^{1}_{\lambda',\lambda}( \theta_h)  \epsilon^{\mu}_{I} (P,\lambda'), 
\end{equation}
with the 
projection matrix $\textbf{H}^{1}_{I,\delta}$ given by 
\begin{widetext}
\begin{equation}
    \label{HWM 1}
\textbf{H}^{1}_{I,\delta} =
\begin{pmatrix}
	H_{1,1} ^{1}(\theta_h) & H_{1,0} ^{1}(\theta_h)&H_{1,-1} ^{1}(\theta_h) \\\\
	H_{0,1}^{1}(\theta_h) & H_{0,0} ^{1} (\theta_h) &H_{0,-1}^{1}(\theta_h)  \\\\
	H_{-1,1}^{1}(\theta_h) & H_{-1,0}^{1} (\theta_h)&H_{-1,-1}^{1}(\theta_h)  
	\end{pmatrix}=\begin{pmatrix}
	\frac{(1+\cos \theta_h)}{2} & -\frac{\sin \theta_h}{\sqrt{2}} &\frac{(1-\cos \theta_h)}{2}\\\\
	\frac{\sin\theta_h}{\sqrt{2}} & \cos\theta_h & -\frac{\sin\theta_h}{\sqrt{2}}\\\\
	\frac{(1-\cos\theta_h)}{2} & \frac{\sin\theta_h}{\sqrt{2}} &\frac{(1+\cos\theta_h)}{2}
	\end{pmatrix}.
	\end{equation} 
\end{widetext}	
Here, the
projection matrix $\textbf{H}^{1}_{I,\delta}$
provides the representation of the interpolating helicities in terms of the IFD Jacob-Wick helicities with $\theta_h = 0$ intrinsically. 
In this respect, 
$H^{1}_{\lambda',\lambda}( \theta_h)$ represents amplitude of  IFD helicity structure related to the $\lambda'$ can be found in the interpolating $ \lambda$ states. 
While we use here the four-component polarization vectors $\epsilon_\delta^\mu (P,\lambda)$ and $\epsilon_I^\mu (P,\lambda')$, namely the spin-1 \((1/2, 1/2)\) chiral representations in Eq.(\ref{summary HWM 1 Pola}), one may express the same equation in the spin-1 
\((0, 1) \oplus (1, 0)\) representation~\cite{PhysRevD.92.105014} using the six-component spin-1 spinors $U_\delta(P,\lambda)$ and $U_I(P,\lambda')$ (see Eq.(\ref{HWM 1 S}) in Appendix \ref{App: C}).
The interpolating Wigner-d matrix components of spin-1 in Eq. (\ref{HWM 1}) can be expressed  in interpolation four vector momenta as
\begin{widetext}
\begin{equation}
\label{HWMP 1}
\begin{split}
H_{1,1} ^{1}(P) =& H_{-1,-1} ^{1}(P) =\frac{(1+\cos \theta_h)}{2}=\frac{1}{2}\big(1+  \frac{\mathbb{P}^2 \cos\delta -P ^{\hat{+}}P_{\hat{-}}\sin\delta}{\mathbb{P} \mathbb{C}|\textbf{P}|}\big),\\
H_{1,-1} ^{1} (P) = &H_{-1,1} ^{1}(P)=\frac{(1-\cos \theta_h)}{2} =\frac{1}{2}\big(1-  \frac{\mathbb{P}^2 \cos\delta -P ^{\hat{+}}P_{\hat{-}}\sin\delta}{\mathbb{P} \mathbb{C}|\textbf{P}|}\big),\\
H_{1,0} ^{1} (P) =& - H_{0,1} ^{1} (P)=H_{0,-1} ^{1}(P) =- H_{-1,0} ^{1} (P)= -\frac{\sin\theta_h}{\sqrt{2}}=\frac{M|\textbf{P}_\perp| \sin\delta}{\sqrt{2}|\textbf{P}|\mathbb{P}},\\
H_{0,0} ^{1} (P) =& \cos\theta_h=\frac{\mathbb{P}^2 \cos\delta -P ^{\hat{+}}P_{\hat{-}}\sin\delta}{\mathbb{P} \mathbb{C}|\textbf{P}|}.
\end{split}
\end{equation}
\end{widetext}
  The profiles of the matrix elements in Eq.(\ref{HWMP 1}) are exhibited in Fig. \ref{FHWM 1} as a function of the interpolation angle $\delta$ and the momentum direction $\theta$ with the use of Eq. (\ref{thetaheq}). For the numerical calculation, we used the same energy and the momentum values taken in Fig. \ref{Fthetah}. As one can see in the Fig.~\ref{FHWM 1}, the profiles in the IFD  ($\delta \rightarrow 0$) ($H^{1}_{\lambda',\lambda} \rightarrow H^{1}_{\lambda',\lambda'}$) exhibit the characteristics ($\theta_h = 0$) of the identity matrix independent of $\theta$ (see Eq. (\ref{HWMP 1})).  
On the other hand, in the LFD ($\delta \rightarrow \pi/4$), $\mathbb{C} \rightarrow 0$, we observe a significant difference in the profiles of the matrix elements $H^{1}_{\lambda',\lambda}$ which can be understood from $\cos \theta_h = \frac{P_v +E_0 \cos \theta}{P_v \cos \theta + E_0}$ and $\sin \theta_h = \frac{M \sin\theta}{P_v \cos \theta + E_0} $ (see Eq.(\ref{thetaheq})). 

While the profiles shown in Fig.
\ref{FHWM 1} between IFD ($\delta=0$) and LFD ($\delta = \pi/4$) vary depending on the momentum direction $\theta$, 
the two particular,  
forward ($\theta =0$) and backward ($\theta = \pi$), moving directions provide distinguished characteristic features in the matrix elements $H^1_{\lambda', \lambda}$. For $\theta = 0$, we have $\theta_h =0$ regardless of $\delta$ and thus see pretty simple constant profiles independent of $\delta$
as the matrix $\textbf{H}^1_{I,\delta}$ in Eq.(\ref{HWM 1}) is just the identity matrix,i.e.
\begin{equation}
\label{identity}
\textbf{H}^1_{I,\delta} =\begin{pmatrix}
	1 & 0 & 0\\\\
	0 & 1 & 0\\\\
	0 & 0 & 1 
	\end{pmatrix} .
\end{equation}
For $\theta = \pi$, however, 
matrix elements $H^1_{\lambda, \lambda}$ are bifurcated by the critical interpolation angle $\delta_c = \tan^{-1}(P_v/E_0)$ as discussed in  illustrating Fig.\ref{Fthetah} with Eq.(\ref{thetaheq1}). 
Here, we see the drastic change of the matrix $\textbf{H}^1_{I,\delta}$ from the identity matrix for 
\( 0 \leq \delta \leq \delta_c \)
to the off-diagonal matrix given by
\begin{equation}
\label{off-diagonal}
\textbf{H}^1_{I,\delta} =\begin{pmatrix}
	0 & 0 & 1\\\\
	0 & -1 & 0\\\\
	1 & 0 & 0
	\end{pmatrix}.
\end{equation}

\vspace{-1cm}

In this context, one may observe similarities between the interpolating spin-1/2 and spin-1 plus and minus states shown in Appendices \ref{App: B} and \ref{App: C}, respectively. However, a significant difference between the spin-1 and spin-1/2 spinor plus states is that the spin-1 state contains a helicity \( 0 \) component, in addition to the usual \( + \) and \( - \) components, which influences not only the structure of the transverse direction component of interpolating  \( 0 \) helicity state but the interpolating plus and minus helicity states as well. This  can be seen in the \( H_{1,0} ^{1} (P)=- H_{0,1} ^{1} (P)=H_{0,-1} ^{1}(P) =- H_{-1,0} ^{1} (P)\) component as  shown in Fig. \ref{FHWM 1}.

The most significant structure can be observed in the component \( H_{0,0}^{1} = \cos(\theta_h) \), which is associated with the interpolating \( 0 \) helicity state decompositions. As explained in the discussion of the \( H_{1,1}^1 \) component, the \( H_{0,0}^1 \) component also has three regions. However, the regions of \( H_{0,0}^1 \) are linked to the structure of the IFD \( 0 \) helicity state in those areas, the value of \( H_{0,0}^1 =1\) . When the conditions \( \delta_c \leq \delta \leq \frac{\pi}{4} \) and \( \theta \rightarrow \pi \) are met, \( H_{0,0}^1 \) changes to \( -1 \), as clearly illustrated in Fig. \ref{FHWM 1}. This change is primarily due to the orientation entanglement of the \( 0 \) helicity state of a spin-1 particle. This type of phase change is not typically observed in the \( 0 \) helicity state of a spin-0 particle. Furthermore, this phase transition occurring at the critical interpolation angle also separates the regions corresponding to IFD from those belonging to LFD.

After analyzing orientation entanglement of spin-1/2 and spin-1 helicity spinors and corresponding  interpolating Wigner-d type of matrices and comparing it with the general Wigner-d matrix of spin ordination, we find that this novel interpolating  Wigner-d matrix of helicity state is valid for any spin-j spinor, more generally, and it can  be written as 
\begin{equation}
	U_{\delta} (P,\lambda)= \sum_{\lambda'=-j}^{j}  H_{\lambda',\lambda} ^{j}  U_{I} (P,\lambda')
    \label{summary HWM general}
	\end{equation}
The derivation of interpolating spin-1/2 and spin-1 spinors can be found in Ref~\cite{PhysRevD.92.105014} and corresponding interpolating Wigner-d expansions are summarized in the Appendix \ref{App: B} and Appendix \ref{App: C}, respectively.

\section{Interpolating Helicity Amplitudes}
		\label{Sec:V}
        
\begin{center}
		\begin{figure}[H]
			\centering	
			\includegraphics[width=7cm]{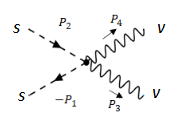}\\
			(b)
			\caption{  Seagull channel Feynman diagram for pair production of spin-1 particles in the pair annihilation of scalar particle and its anti-particle process }
			\label{SStoVV SE}
	\end{figure}
	\end{center}

To exemplify the generalized quantum orientation entanglement in the helicity amplitudes, we examine in this section 
a specific scattering process involving the pair production of two spin-1 (vector) particles. 
In particular, for the simplicity of this work, we focus on the initial pair annihilation of two spinless (scalar) particles in the ``Seagull" contact channel without involving any non-zero impact parameters as depicted in Fig. \ref{SStoVV SE} for the scattering process of $ SS \rightarrow VV $. The process of $ SS \rightarrow VV $ may be described in terms of the Proca~\cite{BENARAB2024129395} spin-1 field $A^\mu$ and the scalar field $\psi$ with its conjugate field $\psi^\dagger$ governed by the following Lagrangian    
\begin{equation}
\label{Lagra}
\begin{split}
\mathcal{L} &= D_\mu \psi (D^\mu \psi)^{\dagger}  -m^2  \psi \psi^{\dagger}  -\frac{1}{4} F_{\mu \nu} F^{\mu \nu} +\frac{1}{2}m_A^2 A_\mu A^\mu ,
\end{split}
\end{equation}
where we denote the covariant derivative as $D^\mu = \partial^\mu +i e A^\mu$ and the field strength tensor as $F^{\mu\nu} =\partial^\mu A^\nu -\partial^\nu A^\mu$. 
Here, the Seagull term may be singled out from this Lagrangian as
\begin{equation}
\label{Seagull-Lagrangian}
\mathcal{L}_{se} = e^2 g^{\mu\nu} A_\mu A_\nu \psi \psi^\dagger.
\end{equation}
For the present work, we look into the orientation entanglement characteristics of the relativistic helicity states 
as close as possible just focusing on the Seagull Lagrangian $\mathcal{L}_{se}$ without involving any other unnecessary complication such as $t$ and $u$ channel terms for the 
$ SS \rightarrow VV $ scattering process.  
The analysis of the full set of channels including $t$-channel, $u$-channel as well as the Seagull channel will be presented separately in the forthcoming work~\cite{Dayananda262}. 
  
Since we are interested in not only "plus" and "minus" helicity states ( transverse polarization vectors), but also zero helicity states of the spin-1 particles ( longitudinal polarization vectors), we specially considered the production of massive spin-1 particles in the corresponding Feynman diagram of Fig.\ref{SStoVV SE}. 
    \vspace{-0.5cm}
    
The interpolating helicity amplitude of the Seagull channel is then given by
\begin{equation}
\begin{split}\label{SStoVV SEeq1}
	M_\delta^{\lambda_1,\lambda_2}=&-2g_{\hat{\mu}{\hat{\nu}}}\epsilon^{*{\hat{\mu}}}(p^3,\lambda_1)\epsilon^{*{\hat{\nu}}}(p^4,\lambda_2)\\
    =&-2g_{\mu\nu}\epsilon^{*{\mu}}_{\delta}(p^3,\lambda_1)\epsilon^{*{\nu}}_{\delta}(p^4,\lambda_2),
    \end{split}
	\end{equation}
where the interpolating basis component of interpolating polarization vectors $\epsilon^{\hat{\mu}}(P,\lambda)= G^{\hat{\mu}} _{\nu }\epsilon^\nu_{\delta} (P,\lambda) = (\epsilon^{\hat{+}} , \epsilon^1 , \epsilon^2 , \epsilon^{\hat{-}} )$ as given by 
Eqs.(\ref{Pola delta2}) and (\ref{HWM 1 Pola}), respectively.
Apparently, the helicities $\lambda_1$ and $\lambda_2$ here denote the interpolating helicities.

As we explained in  Sec.\ref{Sec:IV.B}, each outgoing interpolating polarization vector can now be expanded in the basis of instant form polarization vectors.
When we expand the interpolating polarization vector in terms of the 
instant form (IFD) polarization vectors, the coefficients of the expansion denoted by  $H^1_{\lambda_1',\lambda_1} (p^3)$  and $H^1_{\lambda_2',\lambda_2} (p^4)$ for each outgoing vector particles are multiplied to each other. Consequently, we may rewrite
Eq. (\ref{SStoVV SEeq1}) as	\begin{equation}
	\label{SStoVV SEeq2}
M_\delta^{\lambda_1,\lambda_2}=-2 \sum_{\lambda_1',\lambda_2'}  C_{\lambda_1',\lambda_2'} H^1_{\lambda_1',\lambda_1} (p^3) H^1_{\lambda_2',\lambda_2} (p^4),
	\end{equation}
where 
$C_{\lambda_1',\lambda_2'}=\epsilon_I^{*{\mu}}(p^3,\lambda_1')\epsilon^{*}_{{{I\nu}}}(p^4,\lambda'_2)  $.  Here, we note that the primed helicities 
$\lambda_1^\prime$ and $\lambda_2^\prime$ denote the instant form (IFD) helicities. 
As $\delta \to 0$, i.e. the limit to
the IFD, $M_\delta^{\lambda_1,\lambda_2} \to M_I^{\lambda_1,\lambda_2}$ where 
$H^1_{\lambda_1',\lambda_1} (p^3)
\to H^1_{\lambda_1',\lambda_1'} (p^3) = I$ and $H^1_{\lambda_2',\lambda_2} (p^4)
\to H^1_{\lambda_2',\lambda_2'} (p^4) = I$, recovering the IFD  helicity amplitude as it must:
 \begin{equation}
	 	M_I^{\lambda_1',\lambda_2'}=  -2 \epsilon_I^{*{\mu}}(p^3,\lambda_1')\epsilon^{*}_{{{I\mu}}}(p^4,\lambda'_2).
		\end{equation}
Therefore, we can find interpolating helicity amplitude as the superposition of the instant form helicity amplitudes specifically given by  
         \begin{equation}
		 			\begin{split}
		 				 \label{SStoVV SEeq3}M_\delta^{\lambda_1,\lambda_2}=& \sum_{\lambda_1',\lambda_2'}   H^1_{\lambda_1',\lambda_1} (p^3) H^1_{\lambda_2',\lambda_2} (p^4) M_I^{\lambda_1',\lambda_2'} ,  \\=&
		 				\sum_{\lambda_1',\lambda_2'}  M^{\lambda_1,\lambda_2}_{\lambda_1',\lambda_2'},
		 			\end{split}
		 \end{equation}
where we specify $M^{\lambda_1,\lambda_2}_{\lambda_1',\lambda_2'}$ as defined by $M^{\lambda_1,\lambda_2}_{\lambda_1',\lambda_2'}= H^1_{\lambda_1',\lambda_1} (p^3) H^1_{\lambda_2',\lambda_2} (p^4) M_I^{\lambda_1',\lambda_2'}$. We note here that each coefficient 
$H^1_{\lambda_1',\lambda_1} (p^3) H^1_{\lambda_2',\lambda_2} (p^4)$ corresponding to the IFD helicity amplitude $M_I^{\lambda_1',\lambda_2'}$ depends on the interpolating angle as expected. We may identify $M^{\lambda_1,\lambda_2}_{\lambda_1',\lambda_2'}$ as the projection amplitudes of the seagull process, where we have those upper indices referring to the outgoing spin-1 vectors' interpolating helicity states projecting onto the instant form helicity states given in the lower indices. Namely, these projection amplitudes show how much the corresponding instant form helicity amplitude structures can be seen in the accumulated interpolating helicity amplitude.	

\begin{center}
			\begin{figure}[H]
				\includegraphics[width=7cm]{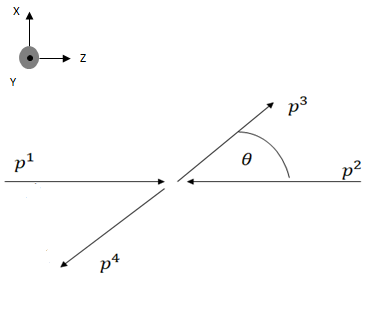}
				\caption{  The center of mass frame (CMF)}
                \label{CMF}
			\end{figure}
		\end{center}

 To make the numerical calculation, we specify the kinematics of the processes. The center of mass frame is chosen to be the initial reference frame as shown in Fig. \ref{CMF}. The moving direction of the initial scalar particles chosen as the $\hat{z}$ and $ -\hat{z}$ directions and two vector particles move in $\theta $ angle ( scattering angle = $\theta $ )  with respect to the z-directions.
		
        \vspace{-1cm}
		Then, the four momenta of the initial and final states particles can be written as
		\begin{equation}
		\begin{split}
		p^1&=(E_0,0,0,P_s),\\
		p^2&=(E_0,0,0,-P_s),\\
		p^3&=(E_0,P_v \sin{\theta},0,P_v\cos{\theta}),\\
		p^4&=(E_0,-P_v \sin{\theta},0,-P_v\cos{\theta}).
		\end{split}
		\end{equation}

Here, $ P_v $ and $ P_s $ are the momenta of vector and scalar particles in the CMF. Then the energy is given by $E_0 =\sqrt{m_s^2 +P_s ^2}=\sqrt{m_v^2 +P_v ^2}$ and  $m_v$  and $m_s$ are the masses of vector and scalar particles respectively.\\
			
After applying the corresponding four vector momenta to the outgoing instant form dynamic polarization vectors, we can find that \begin{equation}
			\label{SeIFDV}
			\begin{split}	
M_I^{\lambda_1',\lambda_2'} = & -2 \epsilon_I^{*{\mu}}(p^3,\lambda_1')\epsilon^{*}_{{{I\mu}}}(p^4,\lambda'_2), \\&=
		\begin{cases}
		2 & \text{if } \lambda_1'=\lambda_2' = \pm, \\
		\frac{-2(E_0 ^2 + P_v ^2)}{ m_v^2}   &  \text{if } \lambda_1'=\lambda_2' = 0, \\
		0  & \text{if } \lambda_1'\neq\lambda_2'.
		\end{cases}
		\end{split}
			\end{equation}
In this process, since the outgoing particles are moving with equal momentum magnitude but in opposite directions, the total orbital angular momentum is zero and also conserved for any scattering angle ($\theta$). Therefore, we can easily understand the conservation of total momentum by considering only the spin-angular momentum conservation. When the spin directions of both particles are parallel to their momentum direction (indicating IFD helicity states of $ ++$), or when the spin directions of both particles are anti-parallel to their momentum directions (indicating IFD helicity states of $--$), the total spin-angular momentum of the system also becomes zero. In both cases, the spin directions of the particles are opposite to each other and effectively cancel each other out. Therefore, in these situations, the total angular momentum of spin-1 states' ($VV$) matches the spin-0 states’ ($SS$) total angular momentum ($=0$), meaning they satisfy the conservation of angular momentum and give the non-zero helicity amplitude.	 Similarly, when both spin-1 states are in 0 spin states (indicating IFD helicity states of $00$), spin angular momentum conservation is satisfied, and we will have non-zero helicity amplitudes. 
But for all the other state combinations,  helicity amplitudes will vanish.
 
Having calculated all the interpolating helicity amplitudes, the results we found were consistent with the well-known symmetry based on parity conservation, that the amplitudes in helicity basis must satisfy~\cite{Bakker,PhysRevD.104.036004}.   Since there are no assigned helicity values for the incoming particles in our $ SS \rightarrow VV $ process, the following relationship among the interpolating helicity amplitudes can be written:
	\begin{equation}
            \label{parity1}
		M_{\delta}^ {-\lambda_1,-\lambda_2}=(-1)^{\lambda_1-\lambda_2}M_{\delta}^{\lambda_1,\lambda_2}.
		\end{equation}
The explicit relations are  $M_{\delta}^{++}=M_{\delta}^{--}$, $M_{\delta}^{+-}=M_{\delta}^{-+}$, $M_{\delta}^{+0}=-(M_{\delta}^{-0})$, $M_{\delta}^{0+}=-(M_{\delta}^{0-})$. 
We also find the corresponding relationship for the projection amplitudes $M^{\lambda_1,\lambda_2}_{\lambda_1',\lambda_2'} $ for our process $SS \to VV$ as given by 
\begin{equation}
             \label{parity2}
		M^ {-\lambda_1,-\lambda_2}_ {-\lambda_1',-\lambda_2'}=(-1)^{\lambda_1-\lambda_2}M^{\lambda_1,\lambda_2}_{\lambda_1',\lambda_2'}.
		\end{equation}

We now analyze each and every independent scattering helicity amplitudes numerically, noting that only the four scattering helicity amplitudes out of the total nine  scattering helicity amplitudes are independent due to the relationship given by Eq.(\ref{parity1}). We may classify those independent helicity amplitudes as 
\begin{center}
\begin{align*}
(\rm \bf A)\;& M_{\delta}^{++}=M_{\delta}^{--}, \\ 
(\rm \bf B)\;& M_{\delta}^{+-}=M_{\delta}^{-+}, \\ 
(\rm \bf C)\;& M_{\delta}^{+0}= -M_{\delta}^{-0}=M_{\delta}^{0+}(\theta+\pi)=-M_{\delta}^{0-}(\theta+\pi), \\ 
(\rm \bf D)\;& M^{00}_\delta.
\end{align*}
\end{center}
For a simple illustration, we take the initial energy of each particle as $2 $ GeV , i.e., $ E_0 = 2$ GeV, and the momenta of scalar particles $P_s = \sqrt{3}$  GeV and vector particles   $P_v = 1$ GeV as we present below. 

\subsection{Interpolating helicity amplitudes: $M_{\delta}^{++}=M_{\delta}^{--}$  }
\label{Amp++}
\begin{widetext}

		\begin{figure}[H]
			\centering
\begin{center}
\begin{tabular}{ |c|c|} 
 		\hline
 		\includegraphics[width=5.5cm]{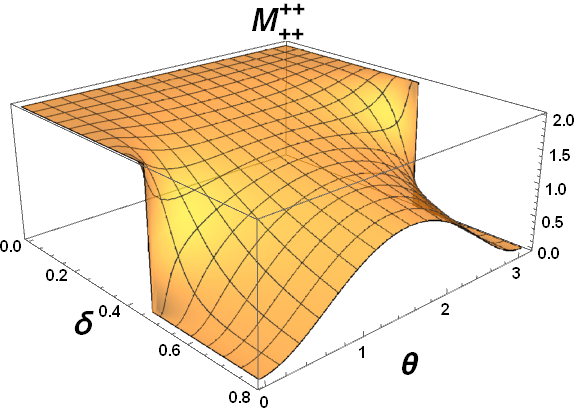}   &  \includegraphics[width=5.5cm]{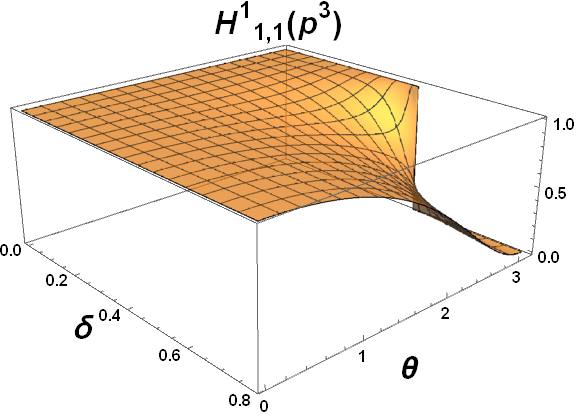}
			\includegraphics[width=5.5cm]{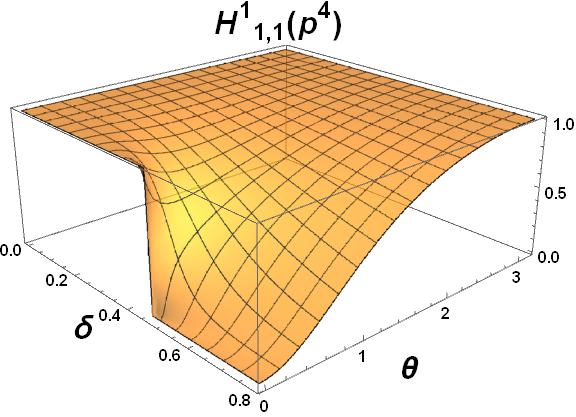} \\[1ex]
            	\hline
 		\includegraphics[width=5.5cm]{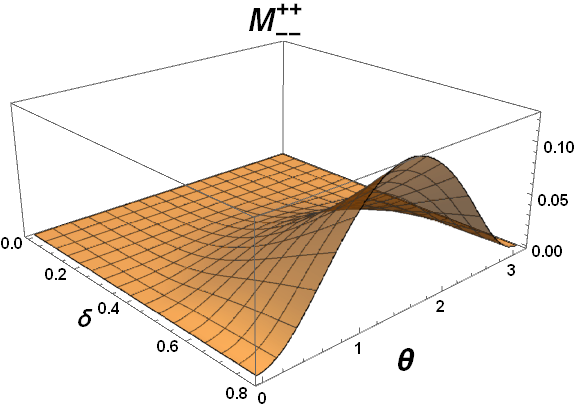} & \includegraphics[width=5.5cm]{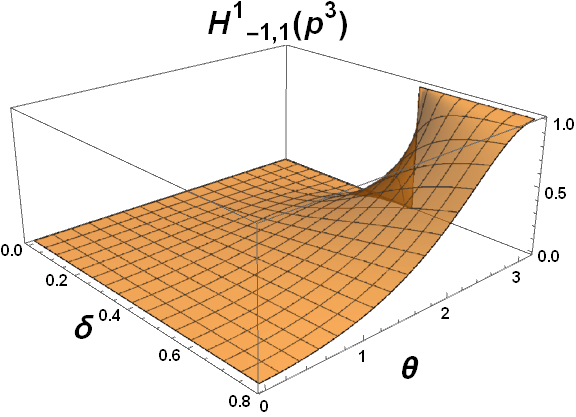}
			\includegraphics[width=5.5cm]{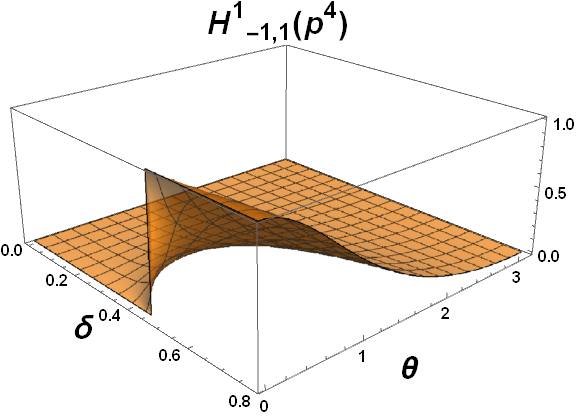} \\[1ex]
            \hline
 	\includegraphics[width=5.5cm]{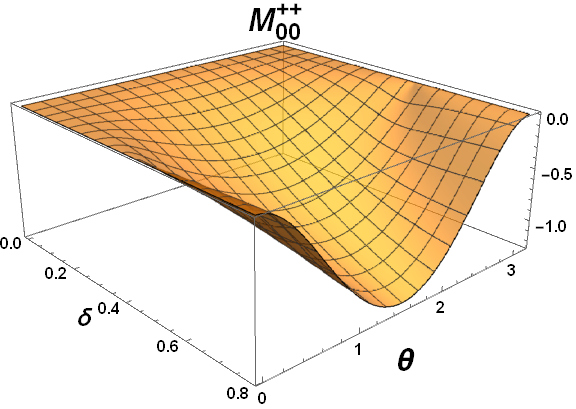} & \includegraphics[width=5.5cm]{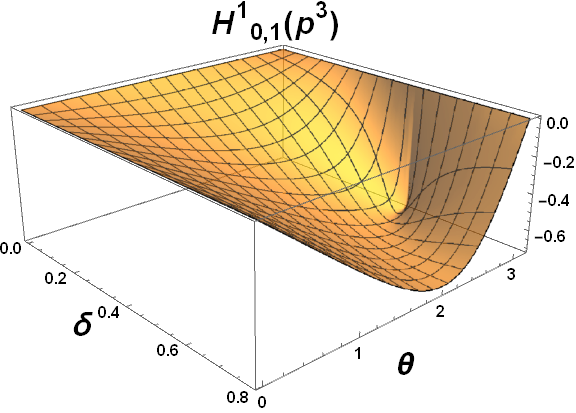} 
			\includegraphics[width=5.5cm]{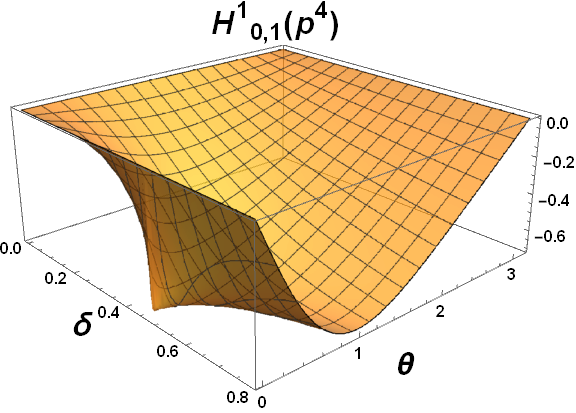}\\[1ex]
 		\hline
 	\end{tabular}
\end{center}
		\centering\caption{Projection helicity amplitudes of $M^{++}$ and the corresponding interpolating Wigner coefficients }
            \label{Mse++AnSum}
		\end{figure}

\end{widetext}

\begin{center}
	\begin{figure}[H]
			\includegraphics[width=7cm]{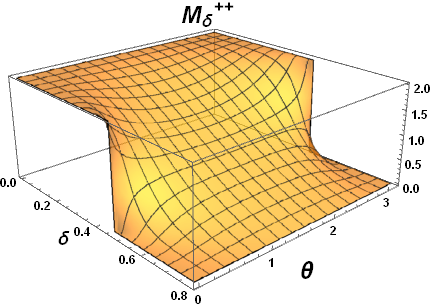}
		\caption{Angular distribution of interpolating helicity amplitude when both helicity values are plus  }
		\label{Mse++An}
	\end{figure}
\end{center}

First, we focus on the angular distribution of the interpolating helicity amplitude, with both outgoing polarization vectors have "plus" interpolating helicity states: $M^{++}_{\delta}$. Using Eqs. (\ref{SStoVV SEeq3}) and (\ref{SeIFDV}), we can write the explicit summation of projection helicity amplitudes corresponds to the  $M^{++}_{\delta}$ as
\begin{equation}
\label{Sumpp}
    M_\delta^{++}=M^{++} _{++} + M^{++} _{--}+ M^{++} _{00},
\end{equation}
where each helicity amplitude projection can be expressed in terms of the interpolating Wigner coefficients that correspond to the two outgoing spin-1 polarization vectors using the same equations,i.e.  
\begin{equation}
\label{projection-helicity-amp}
\begin{split}
   M^{++} _{++} =& 2 (H^1_{1,1} (p^3) H^1_{1,1} (p^4)), \\\\
   M^{++} _{--} =& 2 (H^1_{-1,1} (p^3) H^1_{-1,1} (p^4)),  \\\\
   M^{++} _{00} = & \frac{-2(E_0 ^2 + P_v ^2)}{ m_v^2}  (H^1_{0,1} (p^3) H^1_{0,1} (p^4)).
\end{split}
\end{equation}
The profile of the projection helicity amplitudes, $M^{++} _{++} $, $ M^{++} _{--} $  and $M^{++} _{00}$, 
are plotted in Fig.\ref{Mse++AnSum}
along with their corresponding Wigner d-function components, e.g. $H^1_{1,1}(p^3)$ and $H^1_{1,1}(p^4)$ for $M^{++} _{++}$. All three relations given by Eq.(\ref{projection-helicity-amp}) can be comprehended in Fig.\ref{Mse++AnSum}, e.g. $M^{++} _{++}$ as a product of $H^1_{1,1}(p^3)$ and $H^1_{1,1}(p^4)$ modulo the corresponding factor 2 provided in Eq.(\ref{projection-helicity-amp}) and similarly $M^{++} _{--}$ as well as $M^{++} _{00}$ as their corresponding products as presented in Eq.(\ref{projection-helicity-amp}). Then, the summation of all the projection amplitudes given by Eq. (\ref{Sumpp}) provides the interpolating helicity amplitude  $M^{++}_{\delta}$. The profile of $M^{++}_{\delta}$ helicity amplitude is plotted 
in Fig.~\ref{Mse++An} exhibiting its dependence on the interpolation angle $\delta$ and the scattering angle $\theta$. A few remarks on the profile shown in Fig.~\ref{Mse++An} follow. 

\vspace{-1cm}	

First, as $\delta \rightarrow 0$, we can see the same value we calculated in Eq.(\ref{SeIFDV}) for  $\lambda_1'=\lambda_2'=+ $ in Fig.~\ref{Mse++An} that coincides with the IFD value
$M^{++} _{++}$ shown in the top right corner of the Fig.~\ref{Mse++AnSum} as expected. Note that other projection amplitudes $ M^{++} _{--} $ and $M^{++} _{00}$ in Fig.~\ref{Mse++AnSum} have zero value for $\delta = 0$ due to the conservation of angular momentum.

Next, considering $\theta=0$ and $\theta=\pi$ profiles of $M_{\delta}^{++}$ helicity amplitude, we see clear abrupt changes in the helicity amplitude values.
The profiles of the critical interpolating angles for $\theta=0$ and $\theta=\pi$ 
can be understood from the profiles of $H^1_{1,1}(p^3)$
and $H^1_{1,1}(p^4)$ shown 
in Fig.\ref{Mse++AnSum} which stemmed from the profile $H^1_{1,1}$ exhibited in 
Fig.~\ref{FHWM 1} discussed in Sec.~\ref{Sec:IV.A} identifying the critical interpolation angle $\delta_c$. 
Since we consider the same energy values and the same magnitude of the momenta for both particles with the same helicities, their critical interpolation angles coincide exhibiting the  symmetric helicity profile.

Last but not least, we discuss the limit  
$\delta \to \pi/4$, 
corresponding to LFD, 
where $M^{++}_{\delta \to \pi/4} \to 0$ in Fig.~\ref{Mse++An}.  
While the results for $\theta=0$ and $\theta=\pi$ discussed above illustrate
already $M^{++}_{\delta \to \pi/4} \to 0$ as $\delta  \to \pi/4 > \delta_c$ indicating that the angular momentum conservation in the $SS \to VV$ process prohibits $M^{++}_{\delta > \delta_c}$ for the forward
($\theta=0$) and backward ($\theta=\pi$) scattering. 
For other angles, that's not the criterion. To explain the vanishing helicity amplitude in LFD, we look into the summation of the LF projection amplitudes. For example, if we consider $\theta = \pi/ 2$ angle LF end projection amplitude values,  $M^{++} _{++} (\theta = \pi/ 2)= 1.125 m_e $ $ M^{++} _{--} (\theta = \pi/ 2)=0.125 m_e $, and $M^{++} _{00} (\theta = \pi/ 2) =- 1.25 m_e $. This clearly shows that contribution from $M^{++} _{++}$ is higher even in the LF compare to the  $ M^{++} _{--} $, but the major contribution is gaining from the  $M^{++} _{00}$ amplitude, meaning that the spin structure in the LF actually create more IFD spin-1 "0" helicity structures of both polarization vectors. That value is exactly equal to the magnitude of the summation of    $M^{++} _{++} (\theta = \pi/ 2) $ and $ M^{++} _{--} (\theta = \pi/ 2)$. 

Our discussion here for $M^{++}_\delta$ applies
the same for $M^{--}_\delta$
due to the relation $M^{++}=M^{--}$ given by Eq.(\ref{parity1}). Likewise, the projection amplitudes
work the same, i.e.  $M^{++} _{++} = M^{--} _{--}$, $ M^{++} _{--}= M^{--} _{++}$ and $M^{++} _{00}= M^{--} _{00}$. 

\begin{widetext}
\subsection{Interpolating helicity amplitudes: $M_{\delta}^{+-}=M_{\delta}^{-+}$}
\label{Amp+-}

\begin{center}
		\begin{figure}[H]
\begin{tabular}{ |c|c|} 
 		\hline
 		\includegraphics[width=5.5cm]{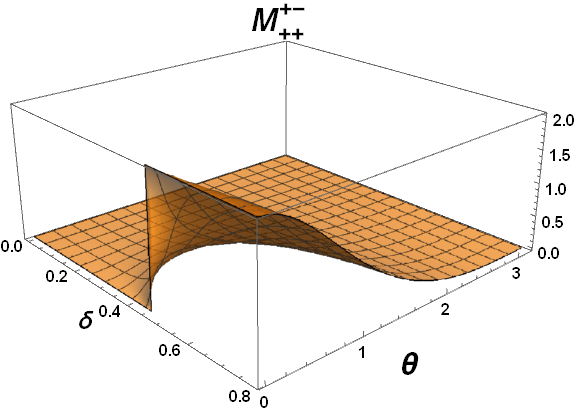}   & \includegraphics[width=5.5cm]{Hpp3.png}
			\includegraphics[width=5.5cm]{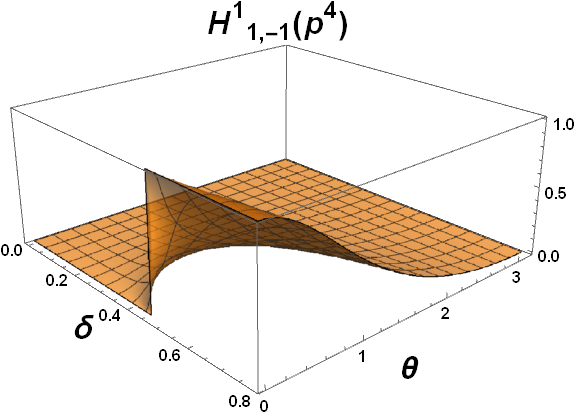} \\[1ex]
            	\hline
 		\includegraphics[width=5.5cm]{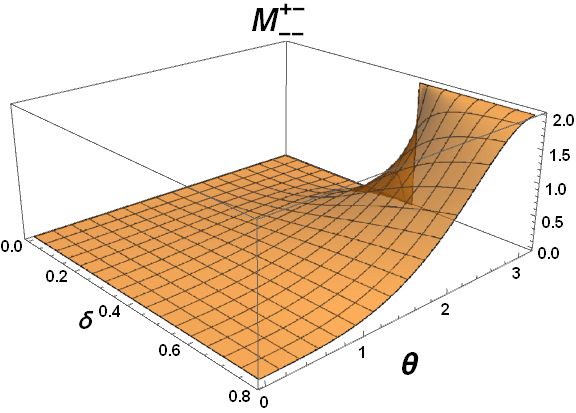}  & \includegraphics[width=5.5cm]{Hmp3.png}
			\includegraphics[width=5.5cm]{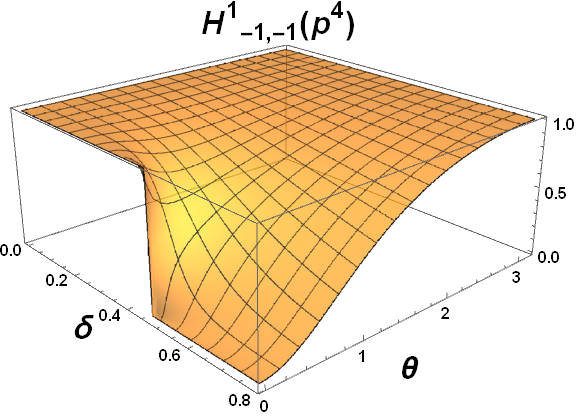} \\[1ex]
            \hline
 	\includegraphics[width=5.5cm]{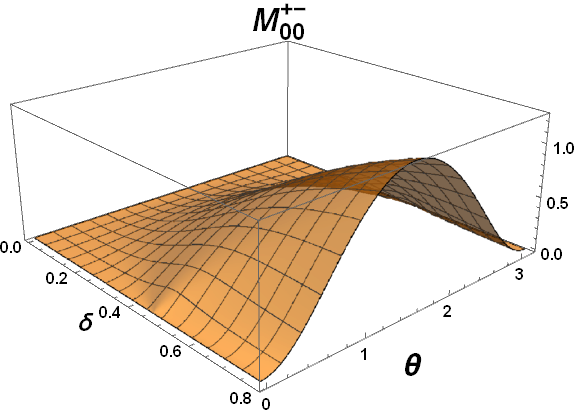} & \includegraphics[width=5.5cm]{Hop3.png} 
			\includegraphics[width=5.5cm]{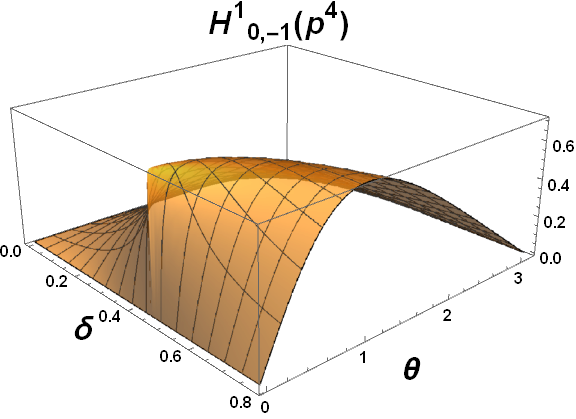}\\[1ex]
 		\hline
 	\end{tabular}
\centering \caption{Projection helicity amplitudes of $M^{+-}$ and the corresponding interpolating Wigner coefficients. }
            \label{Mse+-AnSum}
		\end{figure}
        \end{center}
\end{widetext}
\vspace{-1cm}

We consider here the helicity amplitudes with a combination of plus and minus interpolating helicity amplitudes,  $M^{+-}_{\delta}$ and $M^{-+}_{\delta}$. As 
 $M^{+-}_{\delta} = M^{-+}_{\delta}$ due to Eq.(\ref{parity1}), we won't discuss $M^{-+}_{\delta}$
 here but focus on the amplitude   $M^{+-}_{\delta}$ which
 corresponds to the production of $\epsilon^{{\hat{\nu}}}(p^3,+) $ and $\epsilon^{{\hat{\nu}}}(p^4,-) $ polarization vectors. Using Eqs. (\ref{SStoVV SEeq3}) and (\ref{SeIFDV}) we can write the explicit summation of projection helicity amplitudes corresponds to the $M_{\delta}^{+-}$ amplitude as
 \begin{equation}
 \label{Sumpm}
    M^{+-}_{\delta}=M^{+-} _{++} + M^{+-} _{--}+ M^{+-} _{00},
\end{equation}
where each helicity amplitude projection can also be expressed in terms of the interpolating Wigner coefficients that correspond to the two outgoing spin-1 polarization vectors, namely 
\begin{equation}
\label{projection-helicity-amp+-}
\begin{split}
   M^{+-} _{++} =& 2 (H^1_{1,1} (p^3) H^1_{1,-1} (p^4)), \\\\
   M^{+-} _{--} =& 2 (H^1_{-1,1} (p^3) H^1_{-1,-1} (p^4)),  \\\\
   M^{+-} _{00} = & \frac{-2(E_0 ^2 + P_v ^2)}{ m_v^2}  (H^1_{0,1} (p^3) H^1_{0,-1} (p^4)).
\end{split}
\end{equation}

\begin{center}
 	\begin{figure}[H]
 		\includegraphics[width=7cm]{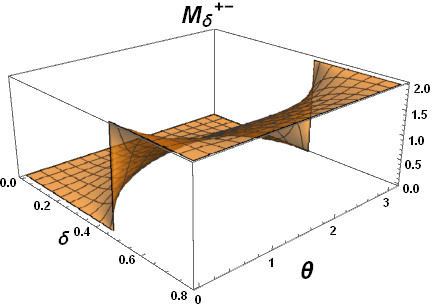}
 		\caption{ Helicity amplitudes when interpolating helicity values are plus and minus  }
 		\label{Mse+-An}
 	\end{figure}
 \end{center}

The profile of the projection helicity amplitudes, $M^{+-} _{++} $, $ M^{+-} _{--} $  and $M^{+-} _{00}$, 
are plotted in Fig.\ref{Mse+-AnSum}
along with their corresponding Wigner d-function components of outgoing polarization vectors $\epsilon^{{\hat{\nu}}}(p^3,+) $ and $\epsilon^{{\hat{\nu}}}(p^4,-) $.  All three relations given by Eq.(\ref{projection-helicity-amp+-}) can be comprehended in Fig.\ref{Mse+-AnSum}, e.g. $M^{+-} _{++}$ as a product of $H^1_{1,1}(p^3)$ and $H^1_{1,-1}(p^4)$ modulo the corresponding factor 2 provided in Eq.(\ref{projection-helicity-amp+-}) and similarly $M^{+-} _{--}$ as well as $M^{+-} _{00}$ as their corresponding products as presented in Eq.(\ref{projection-helicity-amp+-}). Then, the summation of all the projection amplitudes given by Eq. (\ref{Sumpm}) provides the interpolating helicity amplitude  $M^{+-}_{\delta}$. The profile of $M^{+-}_{\delta}$ helicity amplitude is plotted 
in Fig.~\ref{Mse+-An} exhibiting its dependence on the interpolation angle $\delta$ and the scattering angle $\theta$. A few remarks on the profile shown in Fig.~\ref{Mse+-An} follow.

\vspace{-1cm}

First, as $\delta \rightarrow 0$, $M^{+-}_I =0$ since $\lambda_1'\neq \lambda_2'$ (see Eq.(\ref{SeIFDV}))
due to the violation of the angular momentum conservation in $SS \to VV$. Therefore in the IFD, all the projection amplitudes;  $M^{+-} _{++} $, $M^{+-} _{--}$, $M^{+-} _{++} $ vanish and consequently summation of them, $M_{\delta}^{+-}$, vanishes as we can see in Fig.~\ref{Mse+-AnSum} and Fig.~\ref{Mse+-An}, respectively.
Note the violation of the angular momentum conservation as the outgoing vector particles' spins are parallel to each other and add up to be nonzero while the incoming scalar particles are spinless and the Seagull contact interaction doesn't yield any orbital angular momentum. 

Next, it is noticeable that the  $\theta=0$ and $\theta=\pi$ profiles of $M_\delta^{+-}$ in Fig.~\ref{Mse+-An} reveal the critical interpolation angle discussed in Sec.~\ref{Sec:IV.B} (see e.g. Fig.~\ref{FHWM 1}. The helicity amplitude value gets from 0 for 
$0 \le \delta < \delta_c$ to 2 for 
$\delta_c < \delta \le \pi/4$ exhibiting the clear boundary for 
the violation and satisfaction of the angular momentum conservation 
in the $SS \to VV$ scattering process.
In between $\theta=0$ and $\theta=\pi$, i.e. the region of scattering angle  $0 < \theta < \pi$, the highest contribution of $M^{+-} _{00}$ arises near $\theta =\pi/2$ 
and sums up with $M^{+-} _{++}$ and $M^{+-} _{--}$ providing the $M^{+-}_\delta$ profile shown in Fig. 
~\ref{Mse+-An}.  

As mentioned earlier, our discussion
here on $M_\delta^{+-}$ applies the same for $M_\delta^{-+}$ due to the relation $M_\delta^{+-}=M_\delta^{-+}$. Likewise, the projection amplitudes correspond to each other, i.e. $M^{-+} _{++} = M^{+-} _{++}$, $ M^{-+} _{--}= M^{+-} _{--}$ and $M^{-+} _{00}= M^{+-} _{00}$.

\subsection{Interpolating helicity amplitudes: $M_{\delta}^{+0}= -M_{\delta}^{-0}=M_{\delta}^{0+}(\theta+\pi)=-M_{\delta}^{0-}(\theta+\pi)$}
\label{Amp+0}

We discuss here the interpolating helicity amplitudes $M^{+0}_{\delta}$,$M^{0+}_{\delta}$,$ M^{0-}_{\delta}$, and $ M^{-0}_{\delta}$
involving both the interpolating transverse polarization vectors and the interpolating longitudinal polarization vectors. 
While we focus on discussing explicitly $M^{+0} _{\delta}$ here, other amplitudes
$M_\delta^{0+}, M_\delta^{0-}$ and $M_\delta^{-0}$ can be attained by the  relations $M_{\delta}^{+0}= -M_\delta^{-0} = M_{\delta}^{0+}(\theta+\pi)=-M_{\delta}^{0-}(\theta+\pi)$
due to Eq.(\ref{parity1}) as well as the reflection of amplitudes under $\theta \leftrightarrow \theta + \pi$. 

Now, the $M_\delta^{+0}$ helicity amplitude can be written as the summation of the projection helicity amplitudes 
\begin{equation}
\label{Sump0}
    M^{+0}_{\delta}=M^{+0} _{++} + M^{+0} _{--}+ M^{+0} _{00},
\end{equation}
where the helicity amplitude projections are given in terms of the interpolating Wigner coefficient, i.e.
\begin{equation}
\begin{split}
   M^{+0} _{++} =& 2 (H^1_{1,1} (p^3) H^1_{1,0} (p^4)), \\\\
   M^{+0} _{--} =& 2 (H^1_{-1,1} (p^3) H^1_{-1,0} (p^4)),  \\\\
   M^{+0} _{00} = & \frac{-2(E_0 ^2 + P_v ^2)}{ m_v^2}  (H^1_{0,1} (p^3) H^1_{0,0} (p^4)).
\end{split}
\end{equation}

\begin{widetext}
\begin{center}
		\begin{figure}[H]
\begin{tabular}{ |c|c|} 
 		\hline
 		\includegraphics[width=5.5cm]{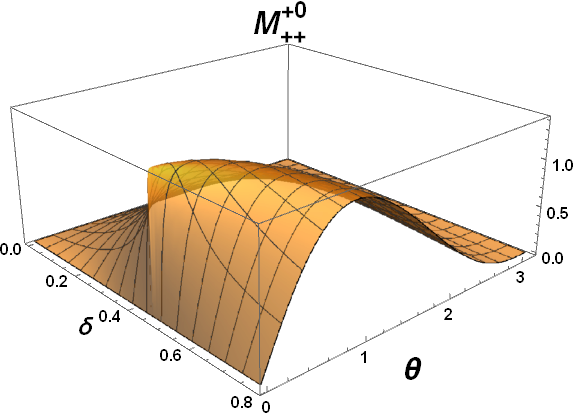}    & \includegraphics[width=5.5cm]{Hpp3.png}
			\includegraphics[width=5.5cm]{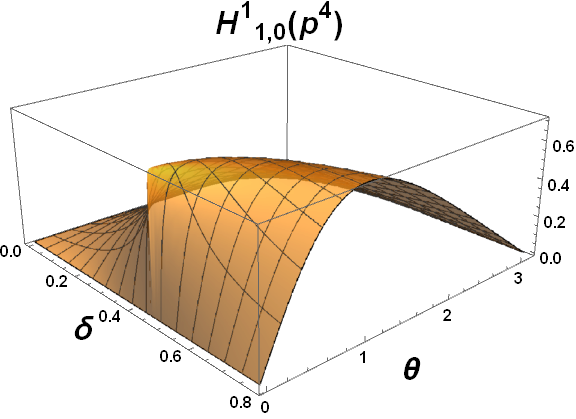} \\[1ex]
            	\hline
 	\includegraphics[width=5.5cm]{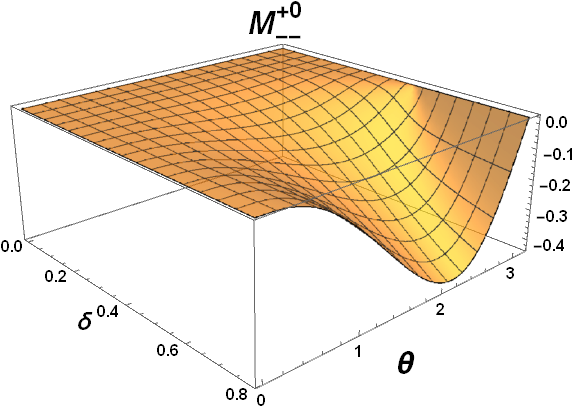}  & \includegraphics[width=5.5cm]{Hmp3.png}
			\includegraphics[width=5.5cm]{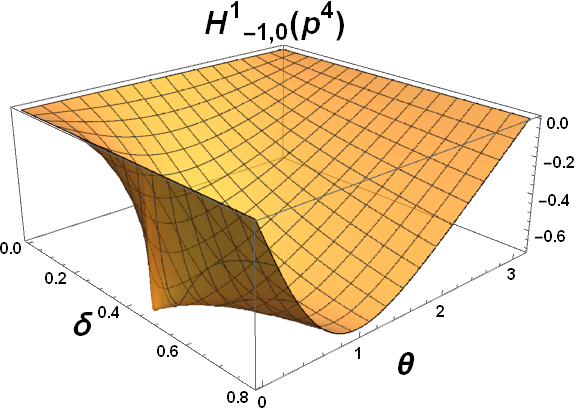}\\[1ex]
            \hline
 	\includegraphics[width=5.5cm]{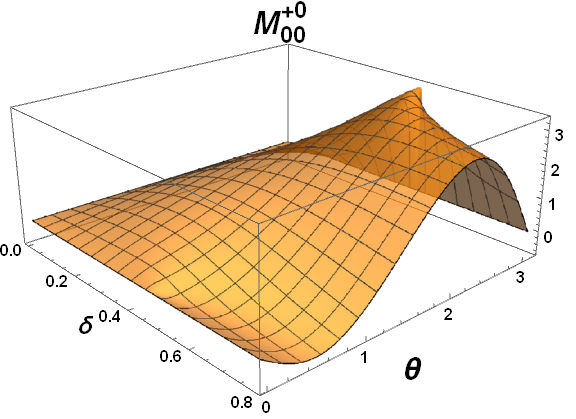} 
			 & \includegraphics[width=5.5cm]{Hop3.png} 
			\includegraphics[width=5.5cm]{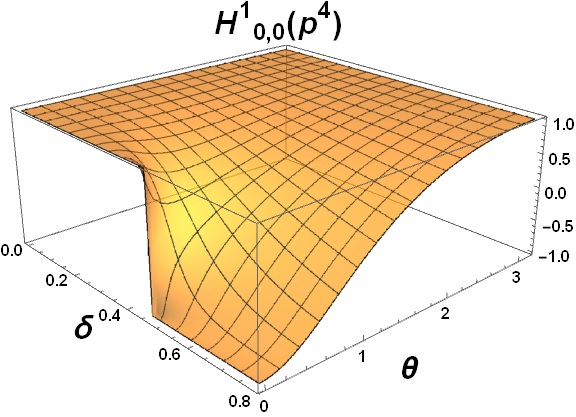}\\[1ex]
 		\hline
 	\end{tabular}
		\centering\caption{Projection helicity amplitudes of $M^{+0}$ and the corresponding interpolating Wigner coefficients. }
            \label{Mse+0AnSum}
		\end{figure}
        \end{center}
\end{widetext}
\vspace{-1cm}
 \begin{center}
	\begin{figure}[H]
		\includegraphics[width=7cm]{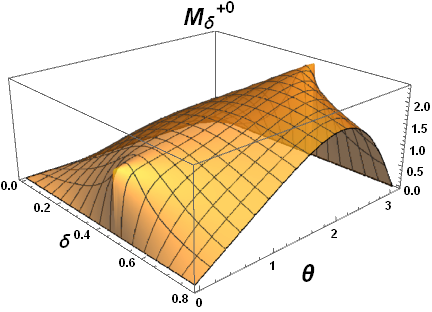}
		\caption{Angular distribution of $M^{+0}$ helicity amplitudes}
		\label{Mse+0An}
	\end{figure}
\end{center}
\vspace{-1cm}

In Fig.\ref{Mse+0AnSum}, the profile of the projection helicity amplitudes, $M^{+0} _{++} $, $ M^{+0} _{--} $  and $M^{+0} _{00}$, are plotted along with their corresponding Wigner d-function components,
$H^1_{1,1}(p^3) H^1_{1,0}(p^4)$, $H^1_{-1,1}(p^3) H^1_{-1,0}(p^4)$ and $H^1_{0,1}(p^3) H^1_{0,0}(p^4)$, respectively. The sum of all the projection amplitudes given by Eq. (\ref{Sump0}) provides the interpolating helicity amplitude $M^{+0}_{\delta}$ as depicted in Fig.~\ref{Mse+0An} exhibiting its dependence on the interpolation angle $\delta$ and the scattering angle $\theta$. 
A few remarks on the profile of $M^{+0}_{\delta}$ shown in Fig.~\ref{Mse+0An} follow. 

As given by Eq.(\ref {SeIFDV}), $\lambda_1'=+$ and $\lambda_2'=0$ yield the vanishing IFD ($\delta = 0$) helicity amplitude 
$M^{+0}_I = 0$ which can be easily understood from the conservation of angular momentum. Not only the projection helicity amplitudes plotted in Fig.~\ref{Mse+0AnSum}
but also their summation exhibited in  Fig.~\ref{Mse+0An} reveal the vanishing IFD helicity amplitude at $\delta=0$. 

Similarly, the vanishing interpolating helicity amplitude $M_\delta^{+0}$ for the forward ($\theta = 0$) and backward ($\theta = \pi$) collinear scattering can be understood from the conservation of the angular momentum regardless of the $\delta$
values ($0 \le \delta \le \pi/4$). Consequently, any sudden change involving the critical interpolation angle $\delta_c$ can be seen neither at $\theta =0$ nor at $\theta = \pi$. Nevertheless, the hint of the critical interpolation angle $\delta_c$ can be observed near $\theta \approx 0$ and $\theta \approx \pi/4$ as somewhat abrupt change exhibited in Fig.~\ref{Mse+0An} for $M_\delta^{+0}$ as well as its projection helicity amplitudes, $M^{+0}_{++},M^{+0}_{--}$ and $M^{+0}_{00}$, displayed in Fig.~\ref{Mse+0AnSum}. It is also interesting to note a nontrivial angular distribution ($\theta$-dependence) of the LFD ($\delta = \pi/4$) helicity amplitude $M^{+0}_{\delta = \pi/4}$ unlike $M^{++}_{\delta = \pi/4}$ and $M^{+-}_{\delta = \pi/4}$ discussed in previous subsections, Sec.\ref{Amp++} and Sec. \ref{Amp+-}, respectively. 

Other interpolating helicity amplitudes $M_\delta^{0+}, M_\delta^{0-}$ and $M_\delta^{-0}$ are related to $M_\delta^{+0}$, namely $M_\delta^{0+}, M_\delta^{0-}$ and $M_\delta^{-0}$, as mentioned earlier. 

\begin{widetext}
\subsection{Interpolating helicity amplitude:  $M^{00}$}
\label{Amp00}

		\begin{figure}[H]
			\centering
\begin{center}
\begin{tabular}{ |c|c|} 
 		\hline
 		\includegraphics[width=5.5cm]{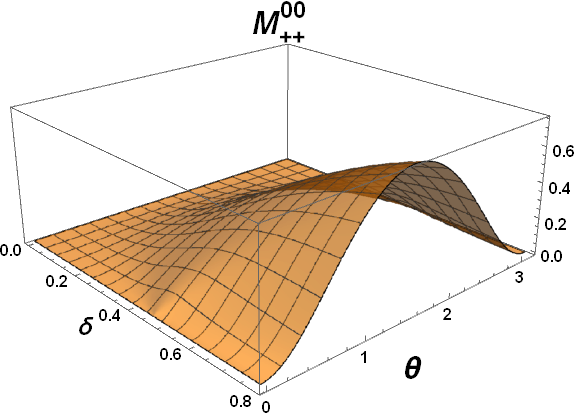}    & \includegraphics[width=5.5cm]{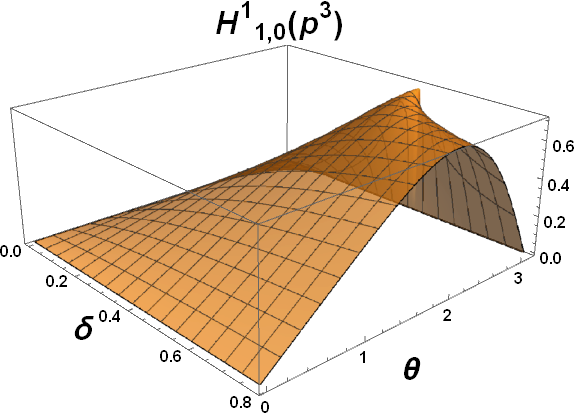}
			\includegraphics[width=5.5cm]{Hpo4.png} \\[1ex]
            	\hline
 	\includegraphics[width=5.5cm]{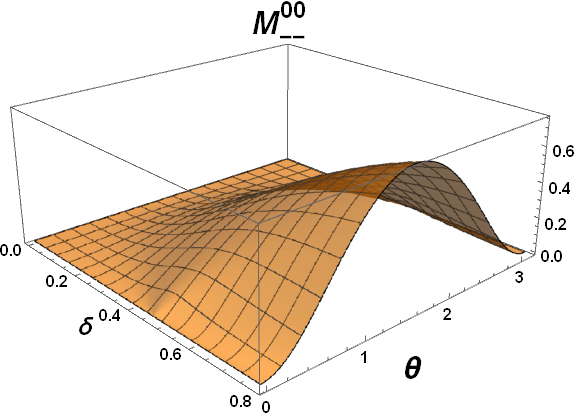}  & \includegraphics[width=5.5cm]{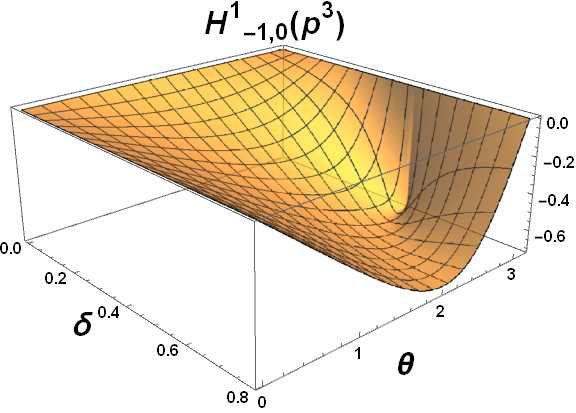}
			\includegraphics[width=5.5cm]{Hmo4.png}\\[1ex]
            \hline
 	\includegraphics[width=5.5cm]{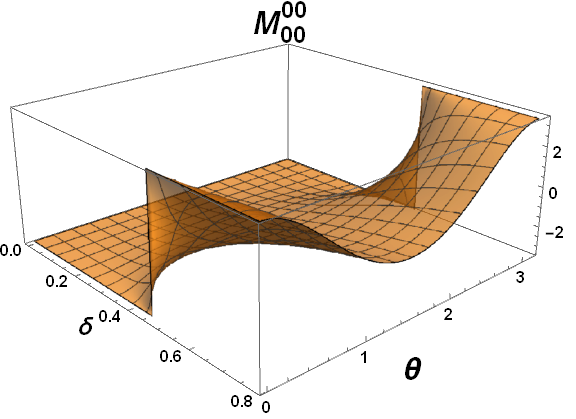} 
			 & \includegraphics[width=5.5cm]{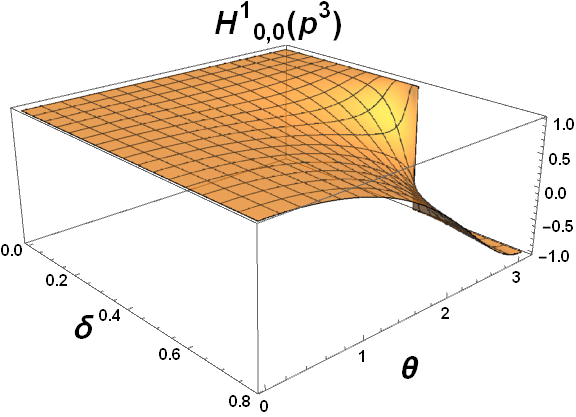} 
			\includegraphics[width=5.5cm]{Hoo4.png}\\[1ex]
 		\hline
 	\end{tabular}
\end{center}
		\centering\caption{Projection helicity amplitudes of $M^{00}$ and the corresponding interpolating Wigner coefficients. }
            \label{Mse00AnSum}.
		\end{figure}
\end{widetext}
\vspace{-2cm}

Having discussed the interpolating helicity amplitudes involving transverse polarizations, e.g. $M_\delta^{++},M_\delta^{+-},M_\delta^{+0}, $, we now discuss the interpolating helicity without involving any transverse polarizations but just the longitudinal polarization, i.e. $M_\delta^{00}$.  
Like the other interpolating helicity amplitudes discussed earlier, $M^{00}_\delta$ can also be written as the summation of projections helicity amplitudes, 
 \begin{equation}
    M^{00}_{\delta}=M^{00} _{++} + M^{00} _{--}+ M^{00} _{00},
\end{equation}
where each helicity amplitude projection can be expressed in terms of the interpolating Wigner coefficients that correspond to the two outgoing spin-1 longitudinal polarization vectors according to Eqs. (\ref{SStoVV SEeq3}) and (\ref{SeIFDV}), i.e.
\begin{equation}
\begin{split}
   M^{00} _{++} =& 2 (H^1_{1,0} (p^3) H^1_{1,0} (p^4)), \\\\
   M^{00} _{--} =& 2 (H^1_{-1,0} (p^3) H^1_{-1,0} (p^4)),  \\\\
   M^{00} _{00} = & \frac{-2(E_0 ^2 + P_v ^2)}{ m_v^2}  (H^1_{0,0} (p^3) H^1_{0,0} (p^4)). 
\end{split}
\end{equation}

\begin{center}
	\begin{figure}[H]
		\includegraphics[width=7cm]{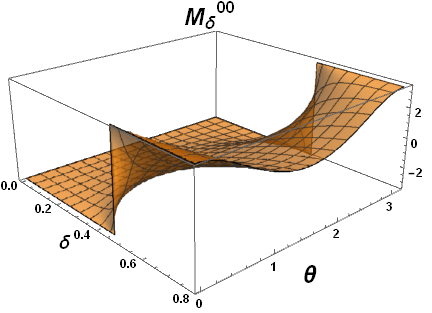}
		\caption{ Helicity amplitudes when both vectors are longitudinal polarization   }
		\label{Mse00An}
	\end{figure}
\end{center}

While the profiles of projection helicity amplitudes, $M^{00} _{++}$,$M^{00}_{--}$ and $M^{00}_{00}$, along with their interpolating Wigner coefficient amplitudes are displayed in the Fig.~\ref{Mse00AnSum}, the profile of interpolating helicity amplitude $M^{00}_\delta$, i.e. the sum of 
$M^{00} _{++}$,$M^{00}_{--}$ and $M^{00}_{00}$, is shown in Fig.~\ref{Mse00An}. 
 
\vspace{-0.7cm}

The IFD helicity amplitude value $M^{00}_{\delta=0}$ can be directly found out using Eq. (\ref{SeIFDV}) for 
$\lambda_1'=\lambda_2'=0 $. Using the numerical values of the energy and momenta for the scattering process $SS \to VV$, i.e. $ E_0 = 2$ GeV, and the momenta of scalar particles $P_s = \sqrt{3}$  GeV and vector particles $P_v = 1$ GeV, 
we get $\frac{-2(E_0^2+P_v^2)}{m_v^2}=-10/3$, and that value can be found in the IFD helicity amplitude of $M^{00}_{\delta=0}$ and its projection amplitude $M^{00}_{00}$ for $\delta = 0$ (IFD) as shown in Fig.~\ref{Mse00An} and the third row of Fig.~\ref{Mse00AnSum}, respectively. 

Examining the $\theta=0$ and $\theta=\pi$ profiles of $M_\delta^{00}$ in Fig.~\ref{Mse00An} as well as its helicity ``00" projection $M^{00}_{00}$ in the third row of Fig.~\ref{Mse00AnSum}, we find the dramatic quantum orientation entanglement effect appearing below ($0 \le \delta < \delta_c$) and above ($\delta_c < \delta \le \pi/4$) the critical interpolation angle $\delta_c$. 
Namely, the IFD helicity amplitude value of $-10/3$
for the IFD branch ($0 \le \delta < \delta_c$)
changes its phase to the LFD 
helicity amplitude value of 
$+10/3$ for the LFD branch 
($\delta_c < \delta \le \pi/4$) and vice versa before/after passing the critical interpolation angle
$\delta_c$. This landscape clearly manifests quantum orientation entanglement of the interpolating helicity amplitude related to the spin-1 ``0" helicity state $|1,0>$, differentiating from the spin-0 ``0" helicity state $|0,0>$. This feature of the quantum orientation entanglement of the spin 1 (vector particle) wavefunction stems from the discussion in Sec.\ref{Sec:II}, where
$R^1_y(\theta_s=\pi)|1,0> = -|1,0>$ was contrasted to 
the rotation invariance of the spin 0 (scalar particle) wavefunction $R^0_y(\theta_s)|0,0>=|0,0>$.
The inheritance of the quantum orientation entanglement for the relativistic spin 1 polarization vector in IFD was then discussed in Sec.\ref{Sec:III}, where we highlighted in particular 
$\epsilon_I^\mu(-P,0) = -\epsilon_I^\mu(P,0)$ under the change of the reference frame $\textbf{P} \to -\textbf{P}$. The relativistic helicity states
interpolating between IFD and LFD were discussed in Sec,\ref{Sec:IV} where 
the bifurcation of the branches of IFD vs. LFD was found with respect to the 
critical interpolation angle $\delta_c$ and the spin orientation by $180^0$ for the given helicity as depicted in Fig.\ref{thetaspi}. 
Thus, the opposite phase between $-10/3$ in the IFD branch and $+10/3$ in the LFD branch in the $\theta=0$ and $\theta=\pi$ profiles of $M_\delta^{00}$ in Fig.~\ref{Mse00An} ( 
as well as $M^{00}_{00}$ in Fig.~\ref{Mse00AnSum})  
stems from the quantum orientation entanglement effect appearing in the spin-1 helicity zero state. 
In contrast, other projection helicity amplitudes $ M^{00}_{++}$ and $ M^{00}_{--}$ in Fig. \ref{Mse00AnSum} trivially vanish for $\theta=0$ and $\theta=\pi$ as they do not have any effect in the  collinear scattering with respect to the interpolating helicity amplitude $M^{00}_\delta$.

In addition, we note that the projection helicity amplitudes 
$M^{00}_{++}$ and 
$M^{00}_{--}$ are equal to each other regardless of all interpolation angles and scattering angles. This symmetry shows that both transverse polarization vectors, whether left-handed or right-handed, have the same effect on the interpolating helicity amplitude with purely longitudinal polarization vectors. Moreover, one can see their larger contribution for the LFD branch in particular around $\theta=\pi/2$ as one may estimate from Eq.(\ref{HWM 1 Pola}). 

Finally, we note that only the $M_I^{00}$ helicity amplitude depends on the momentum of the particle as shown in Eq.~(\ref{SeIFDV}),
explicitly, $M_I^{00}= -2 - 4(P_v^2/m_v^2)$. In contrast to the non-relativistic limit $M_I^{00} \rightarrow -2$ (other non-relativistic results can be seen in Appendix~\ref{App: D}), the Seagull amplitude $M_I^{00} \rightarrow - \infty$ in the higher momentum limit.
We find that in the large momentum limit the divergence of the Seagull amplitude is canceled by 
the contributions from the t-channel and u-channel amplitudes, and thus the total amplitude of the $SS \to VV$ scattering process is finite for the entire kinematic range of the scattering process.   
In Appendix~\ref{App: E}, we discuss how the t and u channel processes contribute to make the total $M_I^{00}$ amplitude finite. We find that the summation of Seagull, t and u channel processes vanish
in the limit $P_v \rightarrow \infty$, equivalently $m_v \to 0$ limit, as one may understand that the longitudinal polarization vectors do not contribute in $m_v \to 0$ limit.

		\section{SUMMARY AND CONCLUSION}
	\label{Sec:VI}

In this work, we analyzed the quantum orientation entanglement stemming from the spin orientation. Starting from the well-known spin-1/2 rotation by $360^0$ for the spin state to acquire the entanglement phase -1, we discussed the $180^0$ rotation of the spin-1 state $|1,0>$ acquiring the phase -1 in contrast to the invariance of the spin-0 state under any rotation. We discussed the Galilean vs. Lorentz transformations of the spin states and explored the relativistic helicity formulations in different forms of dynamics, namely the Jacob-Wick helicity in IFD and the light-front helicity in LFD. We then discussed the interpolating helicity formulation between the Jacob-Wick helicity in IFD and the light-front helicity in LFD.
Analyzing the spin direction of the interpolating helicity states, we identified the critical interpolation angle ($\delta_c$) that bifurcates the IFD branch ($0 \le \delta < \delta_c$) and the LFD branch ($\delta_c < \delta \le \pi/4$) as shown in Figs. \ref{Fthetas} - \ref{FHWM 1}. 

The analysis of interpolating helicity states reveals that they can be expanded in terms of the IFD helicity states as summarized in Eq.(\ref{summary HWM 1 Pola}). The probabilistic coefficients align with the structure of the Wigner-d matrix elements and are related to the angle of interpolating helicity ($\theta_h = \theta_s - \theta$), which is zero for the Jacob-Wick helicity in IFD ($\theta_h = 0$ for $\delta=0$) but represents the relative angle difference between the spin and momentum directions of the interpolating helicity state for $\delta \neq 0$. Our discussion for the expansion of the spin-1 states with the Wigner-d matrix elements can be generalized for any arbitrary spin-j states as summarized in Eq.(\ref{summary HWM general}) and this expansion exhibits distinguished characteristic features of the interpolating helicity states between IFD and LFD. 

To illustrate the quantum orientation entanglement in the interpolating helicity amplitudes, we analyzed the pair production of spin-1 (vector) particles in the
annihilation of two spinless (scalar) particles focusing on their contact interaction, namely the Seagull channel of the scattering process $SS \to VV$ depicted in Fig.\ref{SStoVV SE}. Since the total angular momentum of the initial state in this process is zero, the angular distributions of the interpolating helicity states reveal their orientation entanglement characteristics of the composite system formed by the final state vector particles. We expanded the interpolating Seagull channel helicity amplitudes in terms of the Jacob-Wick helicity amplitudes as given by Eq.(\ref{SStoVV SEeq3}). 
As the Jacob-Wick helicity amplitudes can be understood intuitively ($\theta_h = 0$), 
the expansion in terms of the 
Jacob-Wick helicity amplitudes
allow a systematic extraction of the corresponding projection coefficients associate with the intuitively understandable IFD helicity amplitudes. Analyzing all nine scattering helicity amplitudes
as discussed in Sec.\ref{Sec:V} with \ref{Amp++}, \ref{Amp+-}, \ref{Amp+0} and  \ref{Amp00}, we discussed the bifurcating branch profiles identified by the critical interpolation angle $\delta_c$
in each and every scattering helicity amplitudes. In particular, examining the forward ($\theta = 0$) and backward ($\theta = \pi$) collinear helicity ``00" amplitude $M_\delta^{00}$, we found the dramatic quantum orientation entanglement effect appearing below ($0 \le \delta < \delta_c$) and above ($\delta_c < \delta \le \pi/4$) the critical interpolation angle $\delta_c$. 
Namely, the IFD helicity amplitude value for the IFD branch ($0 \le \delta < \delta_c$)
changes its phase to the LFD 
helicity amplitude value of 
opposite sign for the LFD branch 
($\delta_c < \delta \le \pi/4$) and vice versa before/after passing the critical interpolation angle
$\delta_c$. We attributed this observation of the opposite phase between the IFD branch and the LFD branch in the $\theta=0$ and $\theta=\pi$ profiles of $M_\delta^{00}$ ultimately to the quantum orientation entanglement effect appearing in the spin-1 helicity zero state.

Beyond the contact interaction 
of the Seagull channel analysis, the full set of channels including $t$-channel and $u$-channel for the $SS \to VV$ scattering process as well as its time reversed $VV \to SS$ process deserve further investigation. The work along this line of investigation is underway.  

\section*{Acknowledgment}
This was supported in part by the U.S. Department of Energy (Grant No. DE-FG02-03ER41260). 
The National Energy Research Scientific Computing Center (NERSC) supported by the Office of Science of the U.S. Department of Energy 
under Contract No. DE-AC02-05CH11231 is also acknowledged. 
	
	\appendix
	\section{Interpolating Relations}
	\label{App: A}
	
	The interpolation space time metrics $g^{ \hat{\mu}\hat{\nu} }$ can be written as	\begin{equation}g^{\hat{\mu}\hat{\nu}}= \begin{pmatrix}
	\mathbb{C} & 0 & 0 & \mathbb{S} \\
	0 & -1 & 0 & 0 \\
	0 & 0 & -1 & 0 \\
	\mathbb{S} & 0 & 0 & -\mathbb{C}\\	\end{pmatrix}=g_{\hat{\mu}\hat{\nu}},
	\end{equation} where \, $ \mathbb{C} = \cos{2\delta} $ and  $ \mathbb{S} = \sin{2\delta} $ ,
	thus the covariant and contra-variant indices are related by 
	\begin{equation}
	\begin{split}	a_{\hat{+}}=\mathbb{C}a^{\hat{+}}&+\mathbb{S}a^{\hat{-}} ,\hspace{0.5cm} 	a^{\hat{+}}=\mathbb{C}a_{\hat{+}}+\mathbb{S}a_{\hat{-}}, \\		a_{\hat{-}}=\mathbb{S}a^{\hat{+}}&-\mathbb{C}a^{\hat{-}}\hspace{0.5cm} 	a^{\hat{-}}=\mathbb{S}a_{\hat{+}}+\mathbb{C}a_{\hat{-}}, \\ 
	&a_j= -a^j , (j=1,2).
	\end{split}
	\end{equation}	
	The covariant interpolating space-time coordinates are then easily obtained as 	
	\begin{equation}
	x_{\hat{\mu}}=  g_{\hat{\mu}{\hat{\nu}}} x^{\hat{\nu}}=
	g_{\hat{\mu}{\hat{\nu}}} G^{\hat{\nu}}_\alpha x^{{\alpha}}= R_{\hat{\mu}\alpha}x^{{\alpha}}
	\end{equation}
		
	\begin{equation}
    \label{CovIn}
	R_{\hat{\mu}\alpha}=	g_{\hat{\mu}{\hat{\nu}}} G^{\hat{\nu}}_\alpha= 
	\begin{pmatrix}
	\cos{\delta} & 0 & 0 & -\sin{\delta}\\
	0 & -1 & 0 & 0 \\
	0 & 0 & -1 & 0 \\
	\sin{\delta} & 0 & 0 & \cos{\delta}
	\end{pmatrix}.
	\end{equation}
Here, $R_{\hat{\mu}\nu} (R^{-1})^{\hat{\mu}\alpha}= \delta_\nu ^ \alpha$ . These relations are also apply to the momentum variables.
	Interpolating four-momentum components are shown below
	
	\begin{equation}
	\begin{split}
	P^{\hat{+}}&=(\cos\delta \cosh\beta_3 +\sin\delta \sinh\beta_3)M,\\
	P^{\hat{1}}&= \beta_1 \frac{\sin\alpha}{\alpha}(\sin\delta \cosh\beta_3+\cos\delta \sinh\beta_3 )M,\\
	P^{\hat{2}}&= \beta_2 \frac{\sin\alpha}{\alpha}(\sin\delta \cosh\beta_3+\cos\delta \sinh\beta_3 )M,\\
	P^{\hat{2}}&= \frac{\mathbb{S}P^{\hat{+}}- P_{\hat{-}}}{\mathbb{C}},
	\end{split}	
	\end{equation} where $P_{\hat{-}}=\cos\alpha(\sin\delta\cos\beta_3+ \cos\delta\sinh\beta_3)M$ with $\alpha=\sqrt{\mathbb{C}(\beta_1 ^2 +\beta_2 ^2 )}$ and $M $ is the mass of the particle. The inverted relations of $\beta_1, \beta_2, \beta_3$ and $\alpha$ in terms of the momentum variables are given by 
	\begin{equation}
	\cos\alpha= \frac{ P_{\hat{-}}}{\mathbb{P}} ,\hspace{0.5cm}\sin\alpha = \frac{\sqrt{\bf{P}_{\perp}^2 \mathbb{C} }}{\mathbb{P}},  
	\end{equation}	   
	\begin{equation}
	e^{\beta_3} = \frac{P^{\hat{+}} +\mathbb{P}}{M (\cos\delta +\sin\delta)},
	\end{equation}
	
	\begin{equation}
	e^{-\beta_3} = \frac{P^{\hat{+}} -\mathbb{P}}{M (\cos\delta -\sin\delta)},
	\end{equation}
	\begin{equation}
	\frac{\beta_j}{\alpha}=\frac{P^j}{\sqrt{\bf{P}_{\perp}^2 \mathbb{C} }} \hspace{0.5cm} (j=1,2),
	\end{equation} 
where $\mathbb{P}=\sqrt{(P^{\hat{+}})^2-M^2\mathbb{C}}$ and $ \bf{P}_{\perp} = \sqrt{(P^1)^2 + (P^2)^2} $.

\section{Orientation entanglement of interpolating spin-1/2 spinors }
        \label{App: B}

Interpolating spin-1/2 spinors ~\cite{PhysRevD.92.105014} can also be expanded in the basis of their IFD helicity states using the Wigner-d type of function that depends on the interpolation angle. The relations can be written as

\begin{equation}
\begin{split}
U_{\delta}(P,\frac{1}{2})= &\cos\big(\frac{\theta_h}{2}\big) U_I (P,\frac{1}{2})+ \sin\big(\frac{\theta_h}{2}\big) U_I(P,-\frac{1}{2}),\\
U_{\delta}(P,-\frac{1}{2})= &-\sin\big(\frac{\theta_h}{2}\big) U_I (P,\frac{1}{2})+ \cos\big(\frac{\theta_h}{2}\big) U_I(P,-\frac{1}{2}),
\end{split}
\end{equation}

\begin{center}
	\begin{figure}[H]
		\includegraphics[width=4cm]{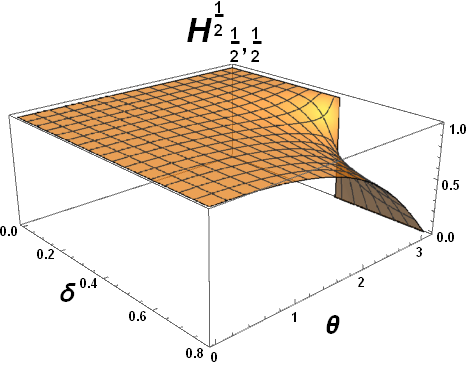}
		\includegraphics[width=4cm]{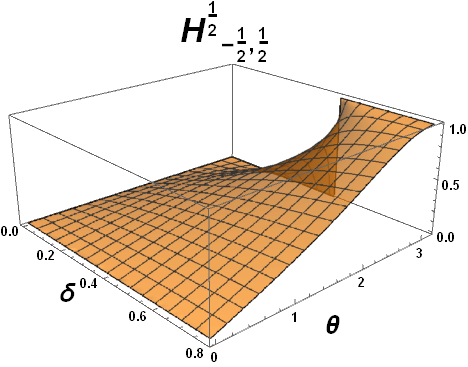}\\
		\centering (a) \hspace{3cm}(b)
		\caption{ Wigner matrix elements of spin-1/2 helicity spinors that depend on the interpolation angle and the particle's moving direction. }
		\label{FHWM 1/2 }
	\end{figure}
\end{center}

Using the above relation, we can summarize the novel interpolating helicity Wigner-d type matrix for spin-1/2 helicity spinors as
\begin{widetext}
\begin{equation}
\label{HWM 1/2}
\mathbf{H}^{1/2}_{I,\delta}(\theta_h) =\begin{pmatrix}
\cos\big(\frac{\theta_h}{2}\big) & -\sin\big(\frac{\theta_h}{2}\big) \\\\
\sin\big(\frac{\theta_h}{2}\big) & \cos\big(\frac{\theta_h}{2}\big)
\end{pmatrix}=\begin{pmatrix}
H_{\frac{1}{2},\frac{1}{2}} ^{\frac{1}{2}}(\theta_h)  & H_{\frac{1}{2},-\frac{1}{2}} ^{\frac{1}{2}} (\theta_h) \\\\
H_{-\frac{1}{2},\frac{1}{2}} ^{\frac{1}{2}}(\theta_h)  & H_{-\frac{1}{2},-\frac{1}{2}} ^{\frac{1}{2}}(\theta_h) 
\end{pmatrix},
\end{equation}    
\end{widetext}
Here, the
	projection matrix $\textbf{H}^{1/2}_{I,\delta} (\theta_h)$
	provides the representation of the interpolating helicities in terms of the IFD Jacob-Wick helicities with $\theta_h = 0$ intrinsically. 
	In this respect, 
	$H^{1/2}_{\lambda',\lambda}( \theta_h)$ represents amplitude of  IFD helicity structure related to the $\lambda'$ can be found in the interpolating $ \lambda$ states.  Since  $\cos\big(\frac{\theta_h}{2})$ and $\sin\big(\frac{\theta_h}{2})$ can be written in the particle's momentum in the interpolating formalism similar to the $\cos\theta_s$ in Eq. (\ref{thetas}) we can write interpolating Wigner matrix elements in terms of particle's interpolating momentum:
\begin{widetext}  
\begin{equation}
\label{HWMP 1/2}
\begin{split}
H_{\frac{1}{2},\frac{1}{2}} ^{\frac{1}{2}}(P)& = H_{-\frac{1}{2},-\frac{1}{2}} ^{\frac{1}{2}}(P) =\sqrt{\frac{1}{2}\big(1+  \frac{\mathbb{P}^2 \cos\delta -P ^{\hat{+}}P_{\hat{-}}\sin\delta}{\mathbb{P} \mathbb{C}|\mathbf{P}|}\big)}=\cos\big(\frac{\theta_h}{2}\big),\\
H_{-\frac{1}{2},\frac{1}{2}} ^{\frac{1}{2}} (P)& =- H_{\frac{1}{2},-\frac{1}{2}} ^{\frac{1}{2}}(P) =\sqrt{\frac{1}{2}\big(1-  \frac{\mathbb{P}^2 \cos\delta -P ^{\hat{+}}P_{\hat{-}}\sin\delta}{\mathbb{P} \mathbb{C}|\mathbf{P}|}\big)}=\sin\big(\frac{\theta_h}{2}\big).\\
\end{split}
\end{equation}
\end{widetext}
The profiles of the matrix elements in Eq.(\ref{HWMP 1/2}) are exhibited in Fig. \ref{FHWM 1/2 } as a function of the interpolation angle $\delta$ and the momentum direction $\theta$ with the use of Eq. (\ref{thetaheq}). For the numerical calculation, we used the same energy and the momentum values taken in Fig. \ref{Fthetah}. As one can see in the Fig.~ \ref{FHWM 1/2 }, the profiles in the IFD  ($\delta \rightarrow 0$) ($H^{1/2}_{\lambda',\lambda} \rightarrow H^{1/2}_{\lambda',\lambda'}$) exhibit the characteristics ($\theta_h = 0$) of the identity matrix independent of $\theta$ (see Eq. (\ref{HWMP 1})).  
On the other hand, in the LFD ($\delta \rightarrow \pi/4$), $\mathbb{C} \rightarrow 0$, we observe a significant difference in the profiles of the matrix elements $H^{1}_{\lambda',\lambda}$ which can be understood from $\cos \theta_h = \frac{P_v +E_0 \cos \theta}{P_v \cos \theta + E_0}$ and $\sin \theta_h = \frac{M \sin\theta}{P_v \cos \theta + E_0} $ (see Eq.(\ref{thetaheq})). 
	
	While the profiles shown in Fig.
	\ref{FHWM 1} between IFD ($\delta=0$) and LFD ($\delta = \pi/4$) vary depending on the momentum direction $\theta$, 
	the two particular,  
	forward ($\theta =0$) and backward ($\theta = \pi$), moving directions provide distinguished characteristic features in the matrix elements $H^{1/2}_{\lambda', \lambda}$. For $\theta = 0$, we have $\theta_h =0$ regardless of $\delta$ and thus see pretty simple constant profiles independent of $\delta$
	as the matrix $\textbf{H}^{1/2}_{I,\delta}$ in Eq.(\ref{HWM 1}) is just the identity matrix,i.e.
    \begin{equation}
	\label{identityH}
	\textbf{H}^{1/2}_{I,\delta} =\begin{pmatrix}
	1 & 0 \\
	0 & 1 \\
	\end{pmatrix} .
	\end{equation} 
    
	For $\theta = \pi$, however, 
	matrix elements $H^{1/2}_{\lambda', \lambda}$ are bifurcated by the critical interpolation angle $\delta_c = \tan^{-1}(P_v/E_0)$ as discussed in  illustrating Fig.\ref{Fthetah} with Eq.(\ref{thetaheq1}). 
	Here, we see the drastic change of the matrix $\textbf{H}^{1/2}_{I,\delta}$ from the identity matrix for 
	\( 0 \leq \delta \leq \delta_c \)
	to the off-diagonal matrix given by
    	\begin{equation}
	\label{off-diagonalH}
	\textbf{H}^{1/2}_{I,\delta} =\begin{pmatrix}
	0 & 1 \\
	-1 & 0 \\
	\end{pmatrix} .
	\end{equation}

\vspace{-1cm}
	
\section{Orientation entanglement of interpolating spin-1 spinors }
    \label{App: C}

The spin-1 polarization vectors $\epsilon^\mu(P,\lambda)$ and corresponding spinors $U(p,\lambda)$ are specifically the chiral \((1/2, 1/2)\) and \((0, 1) \oplus (1, 0)\) representations of the same helicity states~\cite{PhysRevD.92.105014}. While the polarization vectors $\epsilon^\mu(P,\lambda)$ are represented by four-vectors, the corresponding spinors $U(p,\lambda)$ are represented by six-component 
column vectors. This correspondence between the four-component polarization vectors and the six-component spinors is analogous to the correspondence of the four-component gauge field $A^\mu$ with the six-component field strength tensor $F^{\mu\nu}$ in terms of the electric $\vec E$ and magnetic $\vec B$ fields. This correspondence between the polarization vectors and the helicity spinors assures the equivalence between Eq.(\ref{summary HWM 1 Pola}) and Eq.(\ref{summary HWM general}) for $j=1$. 
Thus, the relations in spinor representations are explicitly given by
\begin{widetext}
    
		\begin{equation}
	\label{HWM 1 S}
	\begin{split}
	U_{\delta}(P,+)=&(\frac{1+\cos\theta_h }{2})U_{I}(P,+)+ \frac{\sin\theta_h }{\sqrt{2}}U_{I}(P,0)+(\frac{1-\cos\theta_h }{2})U_{I}(P,-)\\
	U_{\delta}(P,0)=& - \frac{\sin\theta_h }{\sqrt{2}} U_{I}(P,+)+ \cos\theta_h U_{I}(P,0)+\frac{\sin\theta_h }{\sqrt{2}} U_{I}(P,-) \\
	U_{\delta}(P,-)=&(\frac{1-\cos\theta_h }{2})U_{I}(P,+)- \frac{\sin\theta_h }{\sqrt{2}}U_{I}(P,0)+(\frac{1+\cos\theta_h }{2})U_{I}(P,-),
	\end{split}
	\end{equation}
\end{widetext}
where the interpolating spin-1 Wigner-d matrix elements are exactly the same  as discussed in Sec~\ref{Sec:IV.B}.

\section{Non-relativistic helicity amplitudes } \label{App: D}

We summarize here the computation of non-relativistic helicity amplitudes of the $SS\rightarrow VV$ Seagull process
using the non-relativistic polarization vectors in Eq. (\ref{GalvecP}). The non-relativistic scattering helicity amplitudes are then given by
			\begin{equation}
			\label{SeNonRe}
M^{\lambda^1_g,\lambda^2_g}= -2 \epsilon_g^{*{\mu}}(p^3,\lambda^1)\epsilon^{*}_{{{g \mu}}}(p^4,\lambda^2).
			\end{equation}\\
Due to the conservation of angular momentum, only the diagonal ($\lambda_g^1 = \lambda_g^2$) three  helicity amplitudes ($M_{se,g}^{++}$, $M_{se,g}^{--}$ and $M_{se,g}^{00}$) survive as non-zero, while all the off-diagonal ($\lambda_g^1 \neq \lambda_g^2$) six helicity amplitudes
($M_{se,g}^{+-}$,$M_{se,g}^{+0}$, $M_{se,g}^{0+}$,$M_{se,g} ^{0-}$, $M_{se,g}^{-+}$,$M_{se,g}^{-0}$) vanish. Likewise, the off-diagonal six relativistic Jacob-Wick helicity amplitudes in IFD
($M_{se,I}^{+-}$,$M_{se,I}^{+0}$, $M_{se,I}^{0+}$,$M_{se,I} ^{0-}$, $M_{se,I}^{-+}$,$M_{se,I}^{-0}$) vanish.
The three non-vanishing relativistic helicity amplitudes ($M_{se,I}^{++}$, $M_{se,I}^{--}$ and $M_{se,I}^{00}$) can be compared with the non-relativistic helicity amplitudes ($M_{se,g}^{++}$, $M_{se,g}^{--}$ and $M_{se,g}^{00}$). Applying the same kinematics used in Sec.~\ref{Sec:V}, 
we computed both ($M_{se,g}^{++}$, $M_{se,g}^{--}$ and $M_{se,g}^{00}$) and ($M_{se,I}^{++}$, $M_{se,I}^{--}$ and $M_{se,I}^{00}$) and compared to each other. In Fig.~\ref{MseNonRel},
the Non-relativistic Galilean helicity amplitudes and the relativistic IFD helicity amplitudes are denoted by
red and green lines. For the the $SS\rightarrow VV$ Seagull process, the S-wave scattering nature provide the spherical symmetry without any scattering angle $\theta$ dependence in non-vanishing helicity amplitudes.   
Moreover, the Non-relativistic Galilean helicity amplitudes and the relativistic IFD helicity amplitudes for the transverse polarizations coincide to each other, i.e. $M_{se,g}^{++}=M_{se,I}^{++}$, $M_{se,g}^{--}=M_{se,I}^{--}$. 
Thus, the two lines are top on each other and only a single line is shown in (a)$M_{se,g}^{++}=M_{se,I}^{++}$ and (b)$M_{se,g}^{--}=M_{se,I}^{--}$ 
of Fig.~\ref{MseNonRel}. However,
for the longitudinal polarizations, 
the relativistic helicity amplitude gets drastically different from the 
non-relativistic helicity amplitude due to the spacetime mixing of the Lorentz transformation in contrast to the Galilean transformation as discussed in the Sec~\ref{Sec:III}. As shown in Fig.~\ref{MseNonRel}(c), the magnitude of the relativistic helicity amplitude
$M_{se,I}^{00}$ gets larger than the magnitude of the non-relativistic helicity amplitude $M_{se,g}^{00}$. 
While we note the non-relativistic result of $M_{se,g}^{00}
=-M_{se,g}^{++}= -M_{se,g}^{--}$ which reflects the quantum orientation entanglement effect, à la $|1,1>  \rightarrow |1,-1>$, $|1,-1>  \rightarrow |1,1>$, and $|1,0>  \rightarrow -|1,0>$ under 
the rotation of $180^0$, as discussed in Sec.~\ref{Sec:II} and Sec.~\ref{Sec:III}, we see the growth of the relativistic helicity ``00" amplitude $M_{se,I}^{00}$ according to
$M_{se,I}^{00}= -2 - 4(P_v^2/m_v^2)$ given by Eq.~(\ref{SeIFDV}) in  Sec~\ref{Sec:V}. This indicates that the quantum orientation entanglement effect gets enhanced in the relativistic helicity amplitude $M_{se,I}^{00}$ with respect to the non-relativistic helicity amplitude $M_{se,g}^{00}$ as the outgoing vector particle's momentum $P_v$ grows.  
Naively, it alarms though the divergence of $M_{se,I}^{00} \rightarrow - \infty$ as $P_v \to \infty$. However, this alarm in the large $P_v$ limit turns out false due to the contributions beyond the contact interaction, namely, the $t$ and $u$ channel contributions which cancel the divergence from the Seagull channel.  
The physical helicity ``00" amplitude 
summing all channels (Seagull, $t$ and $u$ channels) turn out finite for any $P_v$ values as we explicitly verify in the following Appendix, Appendix~\ref{App: E}.  

\begin{widetext}
	\begin{center}
		\begin{figure}[H]
			\centering
		\includegraphics[width=5.5cm]{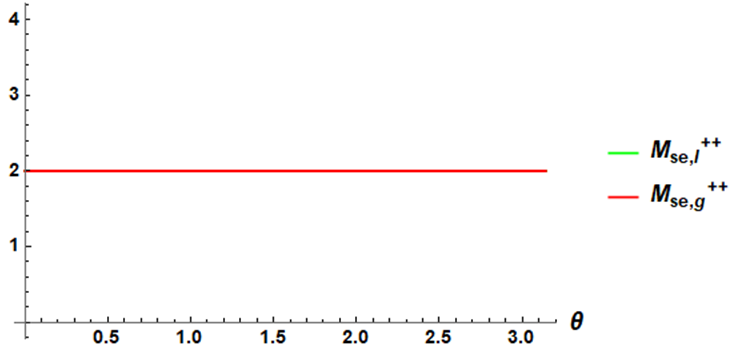} \includegraphics[width=5.5cm]{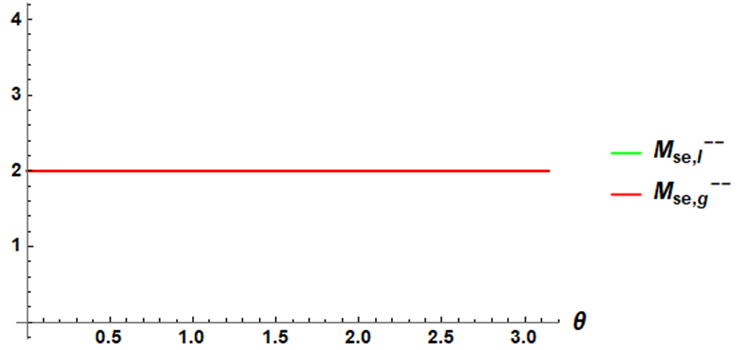}
		\includegraphics[width=5.5cm]{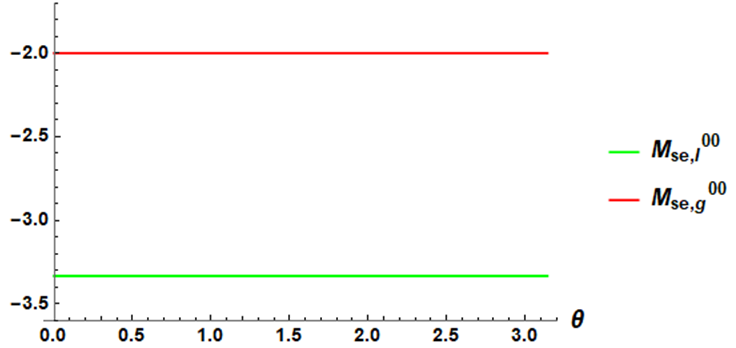}\\ 
         (a)\hspace{5.5cm}(b)\hspace{5.5cm}(c)
        \caption{ Comparing non-relativistic and relativistic helicity amplitudes of $SS\rightarrow VV $ Seagull process (a) ++ , (b) -- (c) 00 }
            \label{MseNonRel}
		\end{figure}
            	\end{center}
                \vspace{-1cm}
\end{widetext}	

     \section{Relativistic effect of longitudinal polarization vectors in the helicity amplitudes } \label{App: E}

\begin{center}
		\begin{figure}[H]
			\centering		\includegraphics[width=5.5cm]{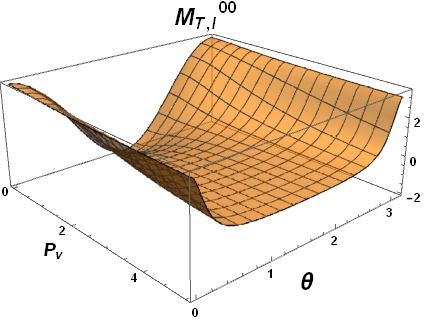} 
 \caption{ The summation of the Seagull, t and u channel helicity
amplitudes }
            \label{MT00Pv}
		\end{figure}
            \end{center}

As alarmed in the previous Appendix, 
Appendix~\ref{App: D}, regarding the divergence of $M_{se,I}^{00}= -2 - 4(P_v^2/m_v^2) \rightarrow - \infty$ 
as $P_v \to \infty$, we computed the $t$ and $u$ channel contributions, $M_{t,I}^{00}$ and $M_{u,I}^{00}$, respectively, beyond the contact interaction of the Seagull channel contribution, $M_{se,I}^{00}$. The results of our computation for $M_{t,I}^{00}$ and $M_{u,I}^{00}$ are given by 
\begin{equation}
\label{Mt00Pv}
    M_{t,I} ^{00}=\frac{4\left(m_v^2 + P_v^2\right)
\left(P_v - \sqrt{-m_s^2 + m_v^2 + P_v^2}\,\cos\theta \right)^2}
{m_v^2 \left(m_v^2 + 2P_v^2
- 2 P_v \sqrt{-m_s^2 + m_v^2 + P_v^2}\,\cos\theta \right)},
\end{equation}
and 
\begin{equation}
\label{Mu00Pv}
M_{u,I} ^{00}=\frac{4\left(m_v^2 + P_v^2\right)
\left(P_v + \sqrt{-m_s^2 + m_v^2 + P_v^2}\,\cos\theta \right)^2}
{m_v^2 \left(m_v^2 + 2P_v^2
+ 2 P_v \sqrt{-m_s^2 + m_v^2 + P_v^2}\,\cos\theta \right)},
\end{equation}
where $m_s$ and $m_v$ are the masses of the scalar and vector particles in the scattering process $SS \to VV$. The forward and backward scattering angle symmetry between $t$ and $u$ channels is apparent from the opposite sign of $\cos \theta$ terms in Eqs.~(\ref{Mt00Pv}) and (\ref{Mu00Pv}). In Fig.~\ref{M00Pv}, the profiles of (a) $M_{se,I}^{00}$, (b) $M_{t,I}^{00}$ and (c) $M_{u,I}^{00}$ are plotted exhibiting the dependence on the outgoing vector particle's momentum ($P_v$) and the scattering angle($\theta$). While $M_{se,I}^{00}$ is independent of $\theta$ representing the S-wave scattering amplitude in contact interaction, $M_{t,I}^{00}$ and $M_{u,I}^{00}$ are dependent on $\theta$ representing the mixture of both S-wave and P-wave scattering amplitudes beyond the contact interaction with non-zero impact parameters. For the forward scattering of $\theta \approx 0$,
$M_{u,I}^{00}$ grows significantly as $P_v$ gets larger while $M_{t,I}^{00}$ remains small almost independent of $P_v$. For the backward scattering of $\theta \approx \pi$, $M_{t,I}^{00}$
grows significantly as $P_v$ gets larger while $M_{u,I}^{00}$ remains small almost independent of $P_v$. In contrast to the negative value of $M_{se,I}^{00}$ which grows as $P_v$ gets larger, the values of $M_{t,I}^{00}$ and $M_{u,I}^{00}$ are positive and grow symmetrically with respect to the forward and backward scattering angles. 
Indeed, the summation of the Seagull, t and u channel helicity amplitudes is found as follow,
 \begin{equation}
 \label{TotalAmp}
 \begin{split}
M_{T,I}^{00}=&     M_{se,I}^{00}+M_{t,I}^{00}+M_{u,I}^{00},\\
=&\frac{2 m_v^2 \left(-2 m_s^2 + m_v^2
+ 2\left(-m_s^2 + m_v^2 + P_v^2\right)\cos(2\theta)\right)}
{\left(m_v^2 + 2 P_v^2\right)^2
- 4 P_v^2 \left(-m_s^2 + m_v^2 + P_v^2\right)\cos^2\theta}.
\end{split}
 \end{equation}
It is evident that not only the divergence in the $P_v \to \infty$ is canceled out in $M_{T,I}^{00}$ but also
Eq.(\ref{TotalAmp}) reveals $M_{T,I}^{00} \rightarrow 0$ in the limit $P_v \rightarrow \infty$ or $M_v \to 0$. The profile of $M_{T,I}^{00}$ is depicted in Fig.~\ref{MT00Pv}. It is significant to note the characteristics of $M_{T,I}^{00} \rightarrow 0$ in the limit $M_v \rightarrow 0$. It indicates that the longitudinal helicity is absent in the massless vector particle as exemplified in the real photon polarization which is entirely transverse. The absence of the longitudinal polarization for the massless vector particle provides 
the reasoning for the absence of 
$M_{T,I}^{00} \to 0$ in the limit 
$M_v \rightarrow 0$. 

\begin{center}
		\begin{figure}[H]
			\centering		\includegraphics[width=7cm]{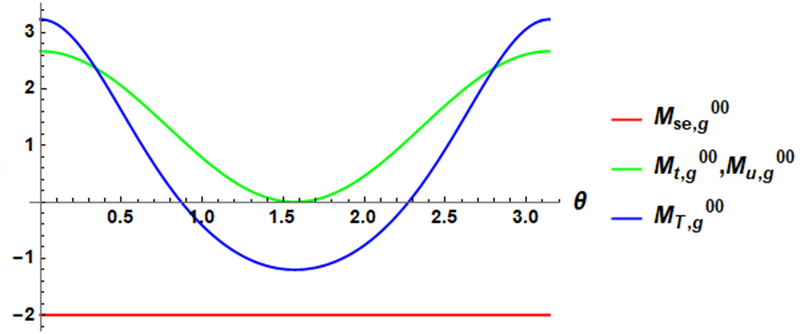} 
 \caption{ Non-relativistic results of Seagull, t and u channel helicity
amplitudes  }
            \label{Mg00F}
		\end{figure}
            \end{center}

In addition, in the limit $P_v \to 0$ one may notice that these relativistic IFD results become identical to the non-relativistic results shown in the Fig.~\ref{Mg00F}. The non-relativsitic amplitudes $M_{se,g}^{00}$, $M_{t,g}^{00}=M_{u,g}^{00}$ and $M_{T,g}^{00}$ are denoted by red, green and blue lines, respectively. Comparing $M_{T,I}^{00}$ in Fig.\ref{MT00Pv} and
$M_{T,g}^{00}$ in Fig.\ref{Mg00F}, we notice that $M_{T,I}^{00}$ at $P_v = 0$
is identical to $M_{g,I}^{00}$ as expected. 
Furthermore, when $m_s = m_v $, the non-relativistic t and u channels vanish at $P_v = 0$ and only the seagull channel survive as one may expect the dominance of the contact interaction in the limit $P_v \to 0$.   

\begin{center}
		\begin{figure}[H]
			\centering	\includegraphics[width=5.5cm]{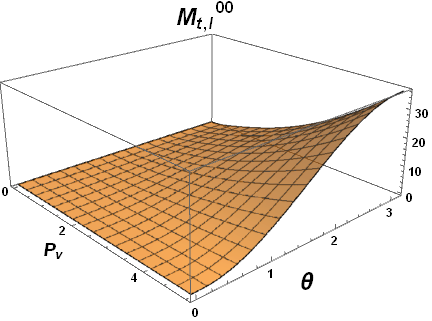} \includegraphics[width=5.5cm]{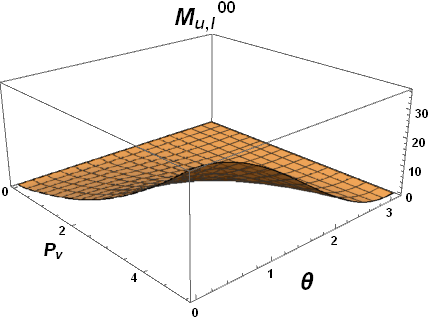}
		\includegraphics[width=5.5cm]{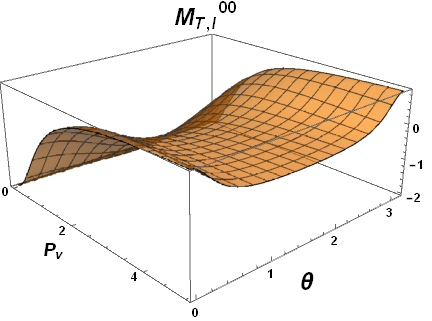}\\ 
         (a)\hspace{5.5cm}(b)\hspace{5.5cm}(c)
        \caption{IFD t(a) u(b) and summation of Seagull,t and u(c) channel processes when outgoing polarization vectors are longitudinal and $m_s=m_v$.}
            \label{M00mv}
		\end{figure}
            	\end{center}

 We can also take the limit $m_s \rightarrow m_v$ for the relativistic helicity amplitudes, where our scalar particle and vector particle have the same momentum as well as the same mass. In that limit, even though the Seagull channel amplitude remains unchanged, the t and u channel helicity amplitudes change. The summation of all three channel helicity amplitudes is shown in Fig.~\ref{M00mv}.

\begin{widetext}
	\begin{center}
		\begin{figure}[H]
			\centering
		\includegraphics[width=5.5cm]{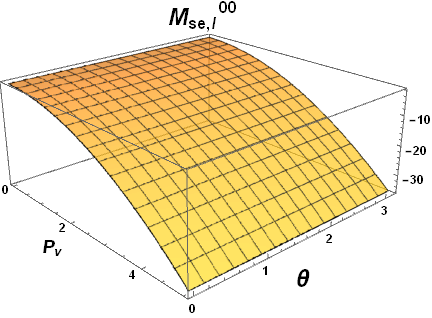} \includegraphics[width=5.5cm]{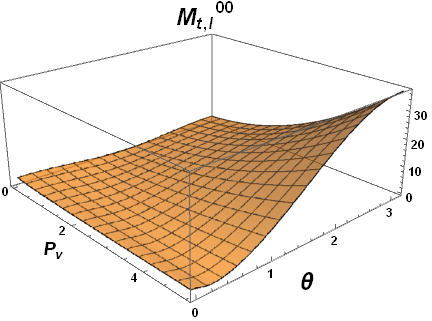}
		\includegraphics[width=5.5cm]{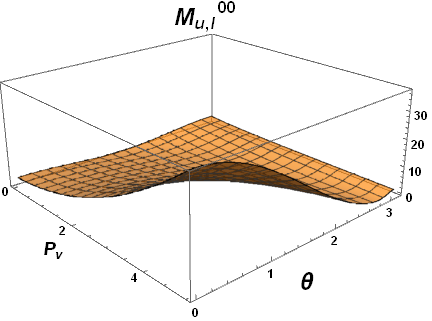}\\ 
         (a)\hspace{5.5cm}(b)\hspace{5.5cm}(c)
        \caption{IFD Seagull(a) t(b) and u(c) channel processes when outgoing polarization vectors are longitudinal}
            \label{M00Pv}
		\end{figure}
            	\end{center}
                \vspace{-1cm}
\end{widetext}

                \vspace{-1cm}

                \vspace{-1cm}

\onecolumngrid
    
                \vspace{-1cm}

\twocolumngrid

\bibliographystyle{apsrev4-2}

\bibliography{bibliography-main}

\end{document}